\documentclass[journal]{IEEEtran}
\IEEEoverridecommandlockouts
\usepackage[utf8]{inputenc}
\usepackage{cite}
\usepackage{floatrow}
\usepackage{titlesec}
\usepackage{amsmath,amssymb,amsfonts,stfloats}
\usepackage{mathtools}
\usepackage{tabularx}
\usepackage[left=0.64in,right=.64in,top=0.75in,bottom=1.1in]{geometry}
\usepackage{algorithmic}
\usepackage{graphicx}
\usepackage{textcomp}
\usepackage{floatrow}
\usepackage{xcolor}
\usepackage{booktabs}

\usepackage{multicol}
\usepackage{acronym}
\usepackage{soul}
\usepackage{url}
\usepackage{mathtools}
\usepackage[font=small]{caption}
\usepackage[caption=false,font=footnotesize]{subfig}
\usepackage{comment}
\def\BibTeX{{\rm B\kern-.05em{\sc i\kern-.025em b}\kern-.08em
    T\kern-.1667em\lower.7ex\hbox{E}\kern-.125emX}}

\newcommand{\linebreakand}{%
  \end{@IEEEauthorhalign}
  \hfill\mbox{}\par
  \mbox{}\hfill\begin{@IEEEauthorhalign}
}

\begin{document}

\title{Enabling NLOS Imaging Capabilities at the Initial Access of 6G Base Stations 
\thanks{The authors are with the Department of Electronics, Information and Bioengineering, Politecnico di Milano, Via Ponzio 24/5, 20133 Milano, Italy. Corresponding author: Davide Tornielli Bellini (email: davide.tornielli@polimi.it).}
}

\author{ Davide~Tornielli~Bellini,~\IEEEmembership{Graduate~Student~Member,~IEEE}, Dario~Tagliaferri,~\IEEEmembership{Member,~IEEE}, Pietro~Grassi,~\IEEEmembership{Graduate~Student~Member,~IEEE},  Davide~Scazzoli,~\IEEEmembership{Member,~IEEE},  
Stefano~Tebaldini,~\IEEEmembership{Senior~Member,~IEEE},
Umberto~Spagnolini,~\IEEEmembership{Senior~Member,~IEEE}
}

\maketitle

\begin{abstract}
Sensing in non-line-of-sight (NLOS) is one of the major challenges for integrated sensing and communication systems. Existing countermeasures for NLOS either use prior knowledge on the environment to characterize all the multiple bounces or deploy anomalous reflectors in the environment to enable communication infrastructure to ''\textit{see behind the corner}''.
This work addresses the integration of monostatic NLOS imaging functionalities into the initial access (IA) procedure of a next generation base station (BS), by means of a non-reconfigurable modular reflector. During standard-compliant IA, the BS sweeps a narrow beam using a pre-defined dedicated codebook to achieve the beam alignment with users. We introduce the imaging functionality by enhancing such codebook with imaging-specific entries that are jointly designed with the angular configuration of the modular reflector to enable high-resolution imaging of a region in NLOS by \textit{coherently} processing all the echoes at the BS. We derive closed-form expressions for the near-field (NF) spatial resolution, as well as for the \textit{effective aperture} (i.e., the portion of the reflector that actively contributes to improve image resolution). 
The problem of imaging of moving targets in NLOS is also addressed, and we propose a maximum-likelihood estimation for target's velocity in NF and related theoretical bound. 
Further, we discuss and quantify the inherent communication-imaging performance trade-offs and related system design challenges through numerical simulations. Finally, the proposed imaging method employing modular reflectors is validated both numerically and experimentally, showing the effectiveness of our concept.

\end{abstract}



\section{Introduction}

Integrated sensing and communication (ISAC) is one of the most promising technologies of the next generation (6G) of wireless networks, combining communication and radar sensing functionalities into a single system, with shared resources \cite{Gonzalez-Prelcic2025}. 
In the broad meaning of sensing, \textit{target localization} is the procedure to estimate the position, velocity, and/or orientation of selected targets of interest (targets' state) from a radar~\cite{10287134}, maximizing the \textit{estimation accuracy}. \textit{Radio imaging}, instead, which is the focus of this paper, refers to the generation of a map of the reflectivity of the environment, from which to infer the number of targets (via detection) and their shape. Rather than estimation accuracy, imaging performance is measured in terms of \textit{resolution}, i.e., the capability of distinguishing closely spaced targets~\cite{manzoni2023wavefield}. 

Research on ISAC covered several areas, considering both \textit{monostatic} (co-located transmitter and sensing receiver) and \textit{bistatic/multistatic} (separate transmitter and sensing receiver) settings. Most of works address target localization, such as \cite{Masouros2025_distributedISAC}, while the integration of imaging into communication systems is more recent \cite{ManzoniCOSMIC,Li2024,11087660,10097213,10540249,IIAC_3D_imaging,zhi2025nearfieldintegratedimagingcommunication}. Some works proposed orthogonal waveform design for single-node ISAC~\cite{ManzoniCOSMIC}, while others consider a monostatic full-duplex base station (BS) that simultaneously serves multiple users (UEs) and images its surroundings through joint beamforming by steering towards fictitious ''virtual UEs''. Works such as \cite{10540249,IIAC_3D_imaging} formulate imaging as an underdetermined inverse scattering problem, where communication signals are used to reconstruct a 2D/3D occupancy map of the environment, whose solution requires \textit{prior information} on targets, e.g,, by imposing (or implicitly assuming) sparsity to regularize the inversion (as they have less measurements than unknowns, i.e., number of pixels). The most recent work along this direction is \cite{zhi2025nearfieldintegratedimagingcommunication}, considering 3D imaging of multiple targets under occlusion effects.

One of the remaining challenges for ISAC is operation in non-line-of-sight (NLOS). On one side, communication in dense multipath and even in pure NLOS is a reality, exploiting reflections from obstacles to enable high-capacity multiple-input-multiple-output (MIMO) links and spatial diversity~\cite{6798744}. In this case, knowing the reflected paths is irrelevant to the data detection, the communication channel is then compensated for at the receiver. Radar sensing, instead, leverages the scattered field from targets, thus \textit{(i)} typical signal processing techniques assume that the forward and backward paths to and from the target are in LOS, and \textit{(ii)} the amount of scattered power from a typical target is much lower than the power of a reflected wave (e.g., by a large wall), thus sensing path-loss is generally higher compared to communication, for a given coverage area. For NLOS operation, therefore, sensing would need an additional reflection from an obstacle with \textit{known position and electromagnetic properties}, at the price of increased path-loss.

\vspace{-0.3cm}\subsection{Related Works on Sensing and ISAC in NLOS}

Sensing in NLOS has been addressed for radars, see for instance \cite{9468353,9553059,9547412,11065141,7362138,8966246,10298635}.
All these works exploit walls and/or a properly placed metallic mirror to overcome the NLOS sensing problem, at the price of being limited in the NLOS exploration angle by specular reflection. Moreover, the imaging resolution and localization accuracy is ultimately dictated by diffraction, and thus by the number of antennas of the radar. A notable exception is \cite{10298635}, where the authors purposely exploit a rough surface to widen the reflection beam beyond diffraction limits, at the price of an average reduction of the signal-to-noise ratio (SNR) compared to use of mirrors. 

Recently, the widespread interest in anomalous mirrors (such as \textit{metasurfaces}) gave a further push in the research for NLOS sensing. Anomalous mirrors are 2D arrays of wavelength-sized unit cells (meta-atoms) whose complex impedance is engineered to accomplish advanced tasks (beam splitting and focusing, anomalous reflection) according to the generalized Snell's law of reflection \cite{7109827}. For dynamic configuration in time, we refer to reconfigurable intelligent surfaces (RISs), while static reflectors are preconfigured during the manufacturing process, yielding orders of magnitude cost reduction compared to RISs at the price of low flexibility. For sensing, metasurfaces enable anomalous reflections beyond specular, obviating the limitations of previous works \cite{9468353,9553059,9547412,11065141,7362138,8966246,10298635} and extending sensing coverage in NLOS.
The vast majority of the literature on metasurfaces for NLOS sensing considers RISs, mainly for localization purposes, see for instance~\cite{Buzzi_RISforradar_journal,10753053}. Here, the authors use a RIS to assist radar operation in NLOS conditions, suggesting its placement close to radar or the target to limit path-loss. To a minor extent, metasurfaces have been explored as a boost for imaging, e.g., see \cite{9299878,jiang2023near,torcolacci2023holographic,10526279,9650562,10618967}. Works like \cite{9299878,jiang2023near,torcolacci2023holographic,10526279}, approach imaging again as an inverse scattering problem---the same principle of ISAC works \cite{10097213,10540249,IIAC_3D_imaging}---requriign prior target information. Differently, works \cite{9650562,10618967} (and references therein) deal with \textit{tomography}, namely imaging the internal structure of targets by processing the waves that propagate through the object in straight lines (only valid at relatively high frequencies). 

%

%

Concerning ISAC-specific works for NLOS operation, the literature is relatively scarce and focused on target localization with existing fifth generation (5G) standard \cite{10694426,jsan13010002,Khosroshahi2024Localization}. A notable exception is the recent work \cite{Paglierani2025_DT_localization}, where digital twins of the propagation environment are used to boost localization in LOS and NLOS conditions with multipath components. The only works dealing with ISAC and imaging in NLOS conditions are our previous preliminary contributions \cite{bellini_2025JCNS,bellini2024multiview}, in which we consider the usage of a BS and a static \cite{bellini_2025JCNS} or reconfigurable \cite{bellini2024multiview} reflector to form an image in NLOS. However, \textit{none} of the existing works address the cost of the integration of imaging functionality into a communication system and related performance trade-offs, which is a crucial aspect for future wireless systems.

\subsection{Contribution}
This paper focuses on the integration of NLOS monostatic imaging in the initial access (IA) procedure of a next-generation BS (\textit{see behind the corner}). During 3GPP standard-compliant IA procedure, the BS employs directive beams (along azimuth) in a periodic beam sweeping procedure, emitting beam-specific signals---the synchronization signal block (SSB)---to achieve the beam alignment of UEs \cite{giordani2018tutorial}. With this procedure, the BS sweeps the entire angular sector (120 deg), thus it is a natural candidate for the integration of imaging functionalities. Sensing with standard compliant signals has been considered in a few works, with either focus on signal selection for ranging ~\cite{1281676,10083170,10200933,10502156,golzadeh2024prs_ambiguity,jopanya2025ssb_uav_detection}, or on innovative beamforming design~\cite{10382696}, with no mention to imaging and NLOS operation. Therefore, we propose to integrate \textit{coherent} NLOS imaging in the IA of the BS by: \textit{(i)} a low-cost pre-configured modular reflector that allows the BS to illuminate and gather the monostatic echoes from targets in a desired region of interest (ROI), exploiting the double reflections; \textit{(ii)} a purposely designed BS beam codebook, jointly designed with the reflector, composed of the standard-compliant one plus additional entries specific for imaging. We owe to a communication-centric system, where the IA is the primary functionality of the BS, and the cost of integrating imaging is represented by the increased latency in the IA due to imaging-specific sweeping. 
The detailed contributions are as follows:
\begin{enumerate}
    \item \textbf{Joint reflector and BS codebook design}. We propose to extend the standard-compliant BS codebook used for IA with additional entries to densify the illumination of the reflector, achieving imaging in the ROI without artifacts (grating lobes) (Sec.~\ref{sec:system_design}). Moreover, we consider a modular angular configuration of the reflector, enhancing imaging resolution and quality w.r.t. BS capabilities. We  derive expressions for the \textit{effective aperture}, i.e., the portion of the reflector determining image resolution, in generic \textit{near-field} (NF) and \textit{far-field} (FF) conditions and as function of design parameters. Useful related insights are derived and discussed. Likewise, we evaluate in closed form the NF spatial resolution (along range and azimuth) as a correction of the well-known FF one. Remarkably, our modular design allows to achieve NF imaging by successive FF acquisitions by the BS, easing the processing. The comparison against fair benchmarks shows the benefits of our proposed imaging system. 
    \item \textbf{Velocity estimation and moving target imaging}. We discuss the effects of target motion on NLOS imaging in NF, showing detrimental image degradation if the target speed is not accounted for in the image generation. Then, we detail a method for maximum likelihood (ML) estimation of the targets' vector velocity in NF using the proposed BS beam sweeping, reporting the theoretical bound and related insights (Sec.~\ref{sec:moving_target_imaging}). 
    \item \textbf{Trade-off and challenges on the integration of NLOS imaging into IA}. We discuss the challenges and trade-offs for imaging integration into IA. In particular, we quantify the bandwidth limitations arising from the usage of SSB signaling during IA, possible countermeasures, and the trade-offs between imaging resolution/SNR/velocity estimation accuracy and IA duration (Sec.~\ref{sec:tradeoffs}). 
    \item \textbf{Experimental testing}. To corroborate the proposed idea, in Sec. \ref{sec:results} we showcase the imaging results of a dedicated in-lab experimental test that demonstrate the feasibility and the benefits of the proposed imaging system with non-reconfigurable modular reflectors.  
\end{enumerate}

We remark the differences between our work and the literature. On one side, ISAC works dealing with imaging \cite{10097213,10540249,IIAC_3D_imaging,zhi2025nearfieldintegratedimagingcommunication} focus on the target reconstruction given communication signals by inverse scattering approaches, but they neither address the cost of the imaging integration as a byproduct, thus ISAC trade-offs, nor consider the NLOS problem. Our work, instead, \textit{(i)} focuses on \textit{direct imaging} that does not require prior target information and \textit{(ii)} explores fundamental communication-imaging trade-offs. Further, we showcase an experimental result.

\section{Imaging Principle} \label{sec:imaging_principle}
\begin{figure}[!t]
\centering
\includegraphics[width=0.9\columnwidth]{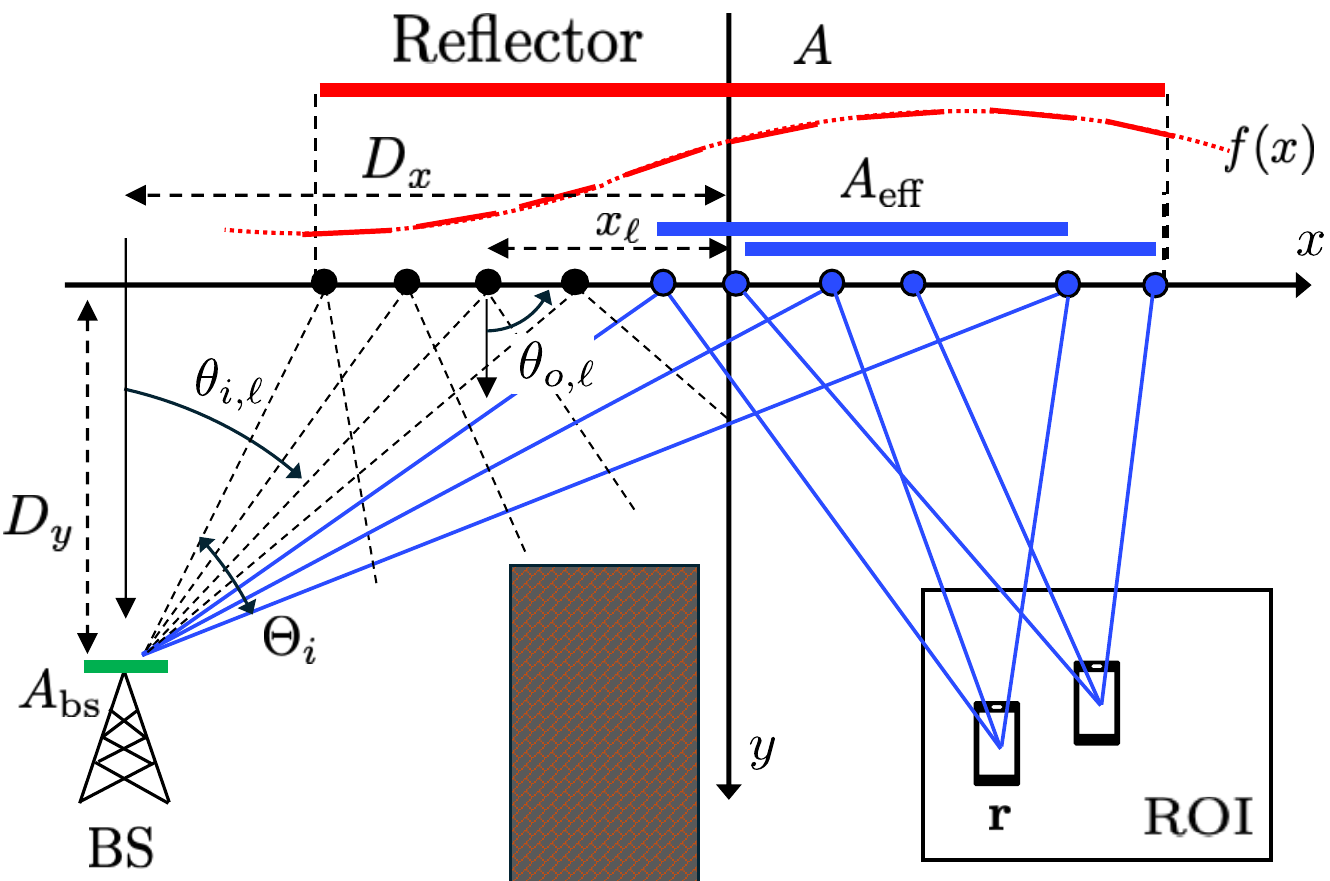}
    \caption{NLOS imaging principle: the reflector allows sensing the targets from multiple ($P>1$) viewpoints (blue lines), increasing the resolution w.r.t. BS. The portion of illuminated reflector that rules the image resolution is denoted as effective aperture $A_{\rm eff}$. }
    \label{fig:enRes}
\end{figure}
Let us consider the system in Fig. \ref{fig:enRes}. We have a reflector of size $A$, located in the origin of the space, that is illuminated by a monostatic BS of aperture $A_{\rm bs}$, located in $\mathbf{x}_{\rm bs} = [-D_x, D_y]^T$. The BS sweeps a pre-defined codebook of $L$ beams, denoted by $\Theta_i$ ($|\Theta_i|=L$), over the reflector; we assume that $\Theta_i$ is dense enough to have a nearly continuous illumination of the reflector, with partially overlapped beams. The reflector is phase-configured such that, for the $\ell$-th beam impinging at angle $\theta_{i,\ell} \in \Theta_i$, we have the reflection angle: 
\begin{equation}\label{eq:angle_periodic_design_varyingincidence}
\begin{split}
        \theta_{o,\ell} = \theta_{i,\ell} + f(x)\big\lvert_{x_\ell=D_y \tan \theta_{i,\ell}-D_x},
\end{split}
\end{equation}
where $f(x)$ is the space-varying angular reflection function, sampled in $x_\ell=D_y \tan \theta_{i,\ell}-D_x$, the coordinate of incidence of the $\ell$-th beam onto the reflector. The angular difference $f(x)$ is designed to image a specific ROI in NLOS, of size $\Delta_x \times \Delta_y$ and centered in $\mathbf{x}_{\rm ROI} \in \mathbb{R}^{2\times 1}$ (details in Section \ref{sec:system_design}). For an arbitrary incidence angle $\theta_i \in \Theta_i$, the reflection angle required to illuminate a static target in $\mathbf{r} = [r_x,r_y]^T$ follows from geometry:
\begin{equation}\label{eq:angle_reflection_target}
\theta_o(\theta_i|\mathbf{x}_{\rm bs},\mathbf{r}) = \arctan\left(\frac{r_x - (D_y \tan \theta_i-D_x)}{r_y}\right).
\end{equation} 
The target will be effectively illuminated only by those beams obtained by finding the $P$ roots of $\theta_{o,\ell} =  \theta_o(\theta_{i,\ell}|\mathbf{x}_{\rm bs},\mathbf{r})$, for $\theta_{i,\ell}\in \Theta_i$. If $P=1$, the target is illuminated from a single point over the plane, thus the resolution is approximately dictated by the BS aperture $A_{\rm bs}$. Conversely, if $P>1$, it follows that the same target is observed by multiple directions, as in Fig.~\ref{fig:enRes} (blue lines), and the spatial resolution may increase if the BS processes the echoes \textit{coherently}. In this latter case, typical of NF operation, the imaging performance is dictated by the \textit{effective aperture} $A_{\rm eff}$, the fraction of the reflector size that contributes to the resolution of the coherent image. The effective aperture is typically larger than the BS aperture, $A_{\rm bs} \leq A_{\rm eff} \leq A$, as detailed in Section \ref{sec:system_design}.  

Noticeably, standard-compliant (3GPP) IA procedures currently used in 5G employ codebooks made of $K$ orthogonal beams (as much as the number of antennas of the BS), herein denoted $\Theta^{\rm 3GPP}_i$, covering a sector of $120$ deg. The spatial sampling $\delta \theta^{\rm 3GPP}_i \simeq 120/K$ [deg] over the reflector is far from being continuous in space, yielding a few observation angles, limiting the imaging resolution and quality. Our method, instead, allows using large pre-configured and low-cost anomalous reflectors that can improve the resolution, at the price of a longer duration of the IA.

\section{System Model}\label{sec:system_model}


Let us refer to the scenario depicted in Fig. \ref{fig:enRes}. The BS is now performing the IA for the UEs in its coverage area while enabling high-resolution imaging of the ROI in NLOS. According to the 3rd generation partnership project (3GPP) standard, during the IA procedure the BS performs a beam sweeping over a pre-defined codebook $\Theta_i^{\rm 3GPP}$ along azimuth, emitting known beam-specific orthogonal frequency division multipleing (OFDM) pilot signal. 
The UEs measure the received signal strength on each of the transmitting (Tx) beams and report the selected (best) beam to the BS during the data transmission phase. To incorporate imaging capabilities within the IA, the BS leverages \textit{(i)} a pre-configured anomalous reflector deployed along the reflection plane (here corresponding to the $x$ axis) that is configured to enable the reflection of the impinging signal from the BS towards the ROI and back to the BS (Section \ref{sec:imaging_principle}), and \textit{(ii)} an additional set $\Theta_i$ of imaging-specific beams to be added to $\Theta_i^{\rm 3GPP}$. In the following, we detail the signal and system model for sensing, which is the peculiarity of this work.

\subsection{Tx Signal}

The BS operates in full-duplex at frequency $f_0$ over a total bandwidth $B$, employing OFDM with $Q$ subcarriers spaced by $\Delta f$. 
The BS is equipped with a full-duplex linear array made of $K$ antennas spaced by $d_{\rm bs}=\lambda_0/2$ along $x$ ($\lambda_0$ being the carrier wavelength), which leads to a physical aperture $A_{\rm bs}=Kd_{\rm bs}$. The reflector is modeled with a linear deployment of $M$ meta-atoms, spaced apart by $d=\lambda_0/4$, each being located in $\mathbf{x}_m = [m d, 0]^T$, $m=-\frac{M}{2}+1,...,\frac{M}{2}$, yielding a reflector size $A=Md$\footnote{The 2D system model outlined in this paper is instrumental to describe the proposed system on the azimuth plane. In practice, planar reflectors deployed on the $xz$ plane are required to guarantee a sufficient SNR. The 3D modeling follows from the present with due adaptations.}. 
The phase configuration of the meta-atoms for imaging is detailed in Section \ref{ssec:ref_design}). 

The Tx signal emitted by the BS on the $\ell$-th Tx beam during the IA is
\begin{equation}\label{eq:TX_signal}
    \mathbf{s}_{\text{RF},\ell}(t) = \sqrt{P_{\rm tx}} \mathbf{f}_\ell \left(\sum_{q=1}^Q S_{q,\ell} \, e^{j 2 \pi q \Delta f t}\right)
    e^{j2\pi f_0 t}, 
\end{equation}
for $\ell = -\frac{L}{2},...,\frac{L}{2}-1$, where $P_{\rm tx}$ is the Tx power, $S_{q,\ell}$ is the unit-energy pilot symbol on the $q$-th subcarrier, dependent on the specific beam, while $\mathbf{f}_\ell = (1/\sqrt{K}) [1,...,e^{- j \pi K \sin \theta_{i,\ell}}]^T \in \mathbb{C}^{K \times 1}$ is the Tx beamforming vector for direction $\theta_{i,\ell}\in \Theta_i$. The duration of the pilot signal, denoted as $\Delta t$, depends on the 3GPP numerology $\mu$ (i.e., on the subcarrier spacing $\Delta f = 15 \times 2^\mu$ [kHz]) and comprises $4$ OFDM symbols. The periodicity of the repetition of the pilot signal is the slot duration $T = 1/2^\mu$ [ms].
 
\subsection{Channel Model and Rx Sensing Signal}

The $\ell$-th beam of the BS impinges the reflector in $\mathbf{p}_\ell = [x_\ell, 0]^T$ and it illuminates a portion determined by the projection of the beam onto the reflector:
\begin{equation}\label{eq:illuminated elements}
    A_{\rm beam,\ell} = M_\ell d \approx \frac{D_y \, \theta_{\rm bs,\ell}}{\sin \theta_{i,\ell} \cos \theta_{i,\ell} }, 
\end{equation}
where $\theta_{\rm bs,\ell} \simeq \lambda_0/(A_{\rm bs} \cos \theta_{i,\ell})$ is the beamwidth of the BS array in direction $\theta_{i,\ell}$, $M_\ell$ being the number of illuminated meta-atoms. The Tx signal $\mathbf{s}_{\text{RF},\ell}(t)$ in \eqref{eq:TX_signal} is directed by the reflector towards the targets in the ROI, it is back-scattered, and it is finally received by the BS after a further reflection by the plane. 
From hereafter, we made the following assumptions: \textit{(i)} the BS uses \textit{narrow beams}, whose projection onto the reflector is less than the reflector size, $A_{\rm beam,\ell} \ll A$; \textit{(ii)} the wavefront of the single beam over the reflector is planar, so the Rx signal by the single beam follows FF model; \textit{(iii)} there are no spatial wideband effects over the illuminated portion $A_{\rm beam,\ell}$. Notice that, although we assume FF for the single beam, the whole reflector can possibly be in the NF of the BS and/or the target. 

Given the monostatic setting, the general model of the received (Rx) sensing signal at the BS, on the $q$-th subcarrier, $\ell$-th beam, $\theta_{i,\ell} \in \Theta_i$, is reported in \eqref{eq:Rx_sig_gen}. 
\begin{figure*}
    \begin{equation}\label{eq:Rx_sig_gen}
    \begin{split}
    Y_{q,\ell} & = \sqrt{P_{\rm tx}} S_{q,\ell} \iint_{\mathbf{x} \in \mathrm{ROI}} \mathbf{f}^H_{\ell}\mathbf{H}^H_{i,\ell}\,  \mathbf{\Phi}_\ell^H \, \mathbf{h}^H_{o,\ell}(\mathbf{x}) \, \Gamma \, \mathbf{h}_{o,\ell}(\mathbf{x}) \, \mathbf{\Phi}_\ell \, \mathbf{H}_{i,\ell} \,  \mathbf{f}_{\ell}\; e^{-j 2 \pi q \Delta f \tau_{\ell}(\mathbf{x})}\, d\mathbf{x} + Z_{q,\ell}\\
    & = \sqrt{P_{\rm tx}} S_{q,\ell} \iint_{\mathbf{x} \in \mathrm{ROI}} \alpha(\mathbf{x}) \beta_\ell(\mathbf{x}) G_\ell(\mathbf{x}) \, e^{-j 2 \pi (f_0 + q \Delta f) \tau_\ell(\mathbf{x})} e^{-j 2 \pi \nu_\ell \ell T}  \, d\mathbf{x} + Z_{q,\ell}
    \end{split}
    \end{equation}
    \hrulefill
\end{figure*}
The Rx signal is the collection of all the echoes from all the scattering locations within the ROI. In \eqref{eq:Rx_sig_gen}, we denote with
\begin{equation}
    \tau_{\ell}(\mathbf{x}) = \frac{ 2\left(\| \mathbf{p}_\ell \hspace{-0.1cm}-\hspace{-0.1cm}\mathbf{x}_{\rm bs} \| \hspace{-0.1cm}+\hspace{-0.1cm} \|\mathbf{x} -\mathbf{p}_\ell \|\right)}{c}= \frac{2\left(D_{i,\ell}\hspace{-0.1cm} + \hspace{-0.1cm}D_{o,\ell}(\mathbf{x})\right)}{c}
\end{equation}
the two-way propagation delay to/from scatterer in $\mathbf{x}$ via double reflection off the incidence point $\mathbf{p}_\ell =[x_\ell, 0]^T = [D_y\tan\theta_{i,\ell}-D_x, 0]^T$. The incidence and reflection channel structures, namely the channel between the BS array to the reflector $\mathbf{H}_{i,\ell} \in \mathbb{C}^{M_\ell \times K}$ and the channel from the reflector to the a target in $\mathbf{x}$, $\mathbf{h}_{o,\ell}(\mathbf{x}) \in \mathbb{C}^{1 \times M_\ell}$, are respectively modelled as 
\begin{align}
    [\mathbf{H}_{i,\ell}]_{m,k} & = \frac{\lambda_0}{4 \pi D_{i,\ell}} e^{-j \frac{2 \pi}{\lambda_0} [D_{i,\ell} + m d \sin \theta_{i,\ell} + k d_{\rm} \sin \theta_{i,\ell}  ]}  \\
    [\mathbf{h}_{o,\ell}(\mathbf{x})]_m & = \frac{\lambda_0 \;e^{j  \pi \nu_\ell(\mathbf{x}) \ell T}}{4 \pi D_{o,\ell}(\mathbf{x})} e^{-j \frac{2 \pi}{\lambda_0} [D_{o,\ell}(\mathbf{x}) + m d \sin \theta_{o,\ell}(\mathbf{x})]}.
\end{align}
These are dependent on the incidence angle $\theta_{i,\ell}$ and the reflection angle $\theta_{o,\ell}(\mathbf{x})$. The reflection channel is also affected by the Doppler shift pertaining to pixel $\mathbf{x}$, $\nu_\ell(\mathbf{x})$, that can be beam-specific due to different observation angles $\theta_{o,\ell}(\mathbf{x})$. The diagonal phase configuration matrix of the illuminated reflector portion is $\mathbf{\Phi}_\ell\in\mathbb{C}^{M_\ell\times M_\ell}$, whose $m$-th diagonal elements are $e^{j\phi_m}$ for $m \in\mathcal{M}_\ell$, the latter being the set of $M_\ell = |\mathcal{M}_\ell|$ meta-atoms illuminated by the $\ell$-th beam. $\Gamma$ is the complex reflection coefficient of the target (that we assume here to be independent on $\ell$, thus for a coherent target). Term $Z_{q,\ell}\sim\mathcal{CN}(0,K\sigma_z^2)$ denotes the additive noise at the BS side, uncorrelated in time and over beams, with power $K \sigma_z^2$ after application of the Rx beamforming (factor $K$).

The second expression in \eqref{eq:Rx_sig_gen} expands the Rx signal obtaining a meaningful expression. The reflectivity of the environment is represented by the space-varying complex scalar $\alpha(\mathbf{x}) = \sqrt{\sigma(\mathbf{x})}e^{j \vartheta(\mathbf{x})}$, where $\sigma$ is the radar cross section of the target, and $\vartheta$ is the scattering phase. The estimation of $\alpha(\mathbf{x})$ is the goal of imaging. The Rx base-band signal is attenuated by the propagation over the double reflection channel, modeled by constant $\beta_\ell(\mathbf{x}) \in \mathbb{R}$:
\begin{equation}
    \beta_\ell(\mathbf{x})= \sqrt{\frac{B \Delta t (4 \pi)^{-7}\lambda_0^6 \,K^4}{ D_{i,\ell}^4 D_{o,\ell}^4(\mathbf{x})}},
\end{equation}
where factor $B \Delta t$ is the time-bandwidth product of the Tx signal and $K^4$ is the BF gain provided by the BS. Although the attenuation of the Tx signal through a double reflection is high, the system harnesses the \textit{reflection gain} in \eqref{eq:Rx_sig_gen} provided by the reflector: 
\begin{equation}
    G_\ell(\mathbf{x}) \hspace{-0.1cm}= \hspace{-0.35cm}\sum_{m,m'\in \mathcal{M}_\ell} \hspace{-0.4cm} e^{j \phi_m}e^{j\phi_{m'}} e^{-j \frac{2\pi d}{\lambda_0}(m+m') \left[ \sin\theta_{i,\ell} -  \sin\theta_{o,\ell}(\mathbf{x})\right]}.
\end{equation} 
The signal to noise ratio (SNR) for the $\ell$-th beam is therefore:
\begin{equation}
    \mathsf{SNR}_\ell(\mathbf{x})= \frac{ P_{\rm tx} B \Delta t \lambda_0^6 \,K^3\, \sigma(\mathbf{x}) |G_\ell(\mathbf{x})|^2}{(4 \pi)^7 D_{i,\ell}^4 D_{o,\ell}^4(\mathbf{x}) \sigma_z^2}.
\end{equation}
As can be seen, in a sensing system relying on double bounces, both reflection gain $|G_\ell|^2$ and beamforming gain $K^3$ are essential to compensate for the severe path loss induced by the forward and backward propagation, in order to have a good SNR both \textit{after} coherent integration over beams (image formation in Section \ref{sec:NLOS_image_generation}), and \textit{before} coherent integration, for the imaging of moving targets (Section \ref{sec:moving_target_imaging}).

\section{Generation of the NLOS Image}\label{sec:NLOS_image_generation}

The image of the ROI is formed by matched filter in space (beams) and frequency (subcarriers). This approach is commonly referred to as \textit{back-projection} in radar literature, and it is a linear imaging method that does not require prior information on targets in the ROI~\cite{manzoni2023wavefield}. For $U$ point targets with position and velocity $\{\mathbf{r}_u,\mathbf{v}_u\}$, the image is:
\begin{equation}\label{eq:BP}
\begin{split}
        I(\mathbf{x},\boldsymbol{\xi}) &= \sum_\ell  \sum_q Y_{q,\ell} \,  \frac{S_{q,\ell}^* \beta^*_\ell(\mathbf{x})  G^*_\ell(\mathbf{x})}{ \sqrt{P_{\rm tx}}|\beta_\ell(\mathbf{x})|^2 |G_\ell(\mathbf{x})|^2} e^{j \varphi_\ell (\mathbf{x},\boldsymbol{\xi})} \\
        & \simeq \sum_{u=1}^U \alpha_u\, \chi\left[\mathbf{x}-\mathbf{r}_u - \delta\mathbf{r}(\mathbf{v}_u-\boldsymbol{\xi})\right] + Z(\mathbf{x})
\end{split}
\end{equation}
where we have denoted the complex conjugate of the propagation phase as
\begin{equation}\label{eq:test_phase}
    \varphi_\ell (\mathbf{x},\boldsymbol{\xi}) = 2 \pi (f_0 + q \Delta f) \tau_\ell(\mathbf{x}) - 2 \pi \nu_\ell(\mathbf{x},\boldsymbol{\xi}) \ell T
\end{equation}
which depends on $D_{i,\ell}$, $D_{o,\ell}(\mathbf{x})$ \textit{and} the Doppler shift of pixel $\mathbf{x}$, moving with apparent speed $\boldsymbol{\xi}$. The back-projection implements the complex conjugate of the propagation model (that we normalize herein to remove path-loss and energy effects) and sum over the subcarriers (the usual matched filtering) and beams (\textit{focusing}). The second term in \eqref{eq:BP} shows the result. The image is an estimation of the reflectivity of the environment, $\{\alpha_u\}$, $u=1,..,U$, but, due to finite aperture and bandwidth, the true response of the environment is convolved with the so-called \textit{spatial ambiguity function} (SAF) $\chi[\mathbf{x}]$.
This is the image of a point target in the considered settings---more details are given in the supplementary material---, whose peak location is in $\mathbf{r}_u + \delta\mathbf{r}(\mathbf{v}_u-\boldsymbol{\xi})$, i.e., the true position of the target plus a velocity-dependent bias $\delta\mathbf{r}(\mathbf{v}_u-\boldsymbol{\xi})$. The latter is a function of the mismatch between the true velocity of the target and the one used to generate the image, $\delta\mathbf{r}(\mathbf{v}_u-\boldsymbol{\xi})\rightarrow 0$ for $\boldsymbol{\xi} \rightarrow \mathbf{v}_u$. 
Finally, term $Z(\mathbf{x})$ denotes the noise in the spatial domain, after image formation. 

\subsection{Spatial Ambiguity Function and Resolution}
Let us express the image in polar coordinates $(R,\psi)$ ($I(\mathbf{x}) \rightarrow I(R,\psi)$), where $R$ is the radius from the center of the reflector and $\psi$ the angle measured from the normal to the reflector. The SAF for a target in $\mathbf{r}_0 = (R_0,\psi_0)$, in generic NF conditions, turns out to be:
\begin{equation}
    \chi[R,\psi] \simeq \mathrm{sinc}\left[\frac{R-R_0}{\rho_R}\right] \mathrm{sinc}\left[\frac{\sin\psi-\sin\psi_0}{\rho_{\psi}}\right]
\end{equation}
where the provided expression is exact in FF. The spatial resolution, i.e., the width of the SAF main lobe, has the expression defined by the following proposition. 

\textit{Proposition}. The range and azimuth resolution in NF can be expressed as \textit{corrections} of well-known FF formulas as follows:
\begin{equation}\label{eq:reoslution_NF}
    \rho^{\rm (NF)}_R = \frac{c}{2 (1\hspace{-0.1cm} + \hspace{-0.1cm}\kappa_R) B},\,\, \rho^{\rm(NF)}_{\psi}=\frac{\lambda_0}{2 (1 \hspace{-0.1cm}-\hspace{-0.1cm}\kappa_\psi) A_{\rm eff}  \cos \psi_0}
\end{equation}
where $A_{\rm eff}$ is the effective aperture on the reflector, introduced in Section \ref{sec:imaging_principle}, and 
\begin{equation}
    \kappa_R  = \frac{f_0}{B} \left[1 -\cos\left(\mathsf{F}_+\right)\right]   \label{eq:range_factor_NF}
\end{equation}
\begin{equation}
    \kappa_\psi = 1 - \frac{2 R_0}{A_{\rm eff} \cos \psi_0} \left[\sin\left(\mathsf{F}_+\right) - \sin\left(\mathsf{F}_-\right) \right] \label{eq:azimuth_factor_NF}
\end{equation}
are the correction factors for the bandwidth $B$ and for the effective area $A_{\rm eff}$, respectively, in which we define the constant
\begin{equation}
    \mathsf{F}_\pm \triangleq \arctan\left(\frac{R_0 \sin\psi_0 \pm A_{\rm eff}/2}{R_0 \cos \psi_0}\right)-\psi_0,
\end{equation}
depending on the position of the target w.r.t. the reflector and on the effective area. FF expressions follow from \eqref{eq:reoslution_NF} by letting $\kappa_R=0$ and $\kappa_\psi=0$. Formulas 

\textit{Proof}. See the supplementary material.

In \eqref{eq:reoslution_NF}, we have $\kappa_R \geq 0$, meaning that NF always increases range resolution, as expected. In FF $A_{\rm eff}/R_0 \rightarrow 0$ and so does $\mathsf{F}_+$, which yields $\kappa_R \rightarrow 0$. In fact, whenever there is no curvature of the wavefront, the resolution is fully determined by the bandwidth $B$. Concerning azimuth resolution, instead, we have $0 \leq \kappa_\psi \leq 1$ for $\psi_0 \leq 35$ deg and $\kappa_\psi < 0$ for $\psi_0 > 35$ deg. This can be shown by expanding $\rho^{\rm(NF)}_{\psi}$ at the first order around $A_{\rm eff}/R_0 \rightarrow 0$, not reported for brevity.  
Expressions of NF range resolution have been derived in the literature for some selected array configurations\cite{wachowiak2025approximationrangeambiguityfunction}. Here, we obtain simple yet effective expressions that hold for monostatic sensing systems, and outline the relations with the FF formulas. 

\subsection{SNR}
The SNR of the coherent image $I(\mathbf{x},\boldsymbol{\xi})$ for the $u$-th target is typically evaluated at the peak of the image ($\mathbf{r}_u$), as it is enhanced by the number of \textit{effective beams} illuminating the target, $L_{\rm eff}$, that correspond to the effective aperture $A_{\rm eff}$. The SNR is therefore:
\begin{equation}\label{eq:SNR_coherent}
    \mathsf{SNR}_u \approx L_{\rm eff} \mathsf{SNR}_{u,0} = \frac{ L_{\rm eff} P_{\rm tx} B \Delta t \lambda_0^6 \,K^3\, \sigma_{u} |G_0|^2}{(4 \pi)^7 D_{i,0}^4 D_{o,0}^4 \sigma_z^2}
\end{equation}
where $|G_0|$ is the reflection gain provided by the $\ell=0$-th beam (illuminating the reflector in the center), and $D_{i,0} = \sqrt{D_x^2 +D_y^2}, D_{o,0} = R_0$ denote the forward and backward distances for the central beam ($\ell=0$). The impact of BS beamforming on the SNR scales with $K^3$, while the reflection gain goes with $|G_0|^2 \propto |M_0|^4 \propto K^{-4}$. Therefore, it would be possible to use a large beam illuminating the whole reflector in a single snapshot, and that would be a SNR-maximizing solution. However, a single snapshot would not allow for velocity estimation, and it is not considered relevant here.

\begin{figure}[!t]
\centering
    \includegraphics[width=\columnwidth]{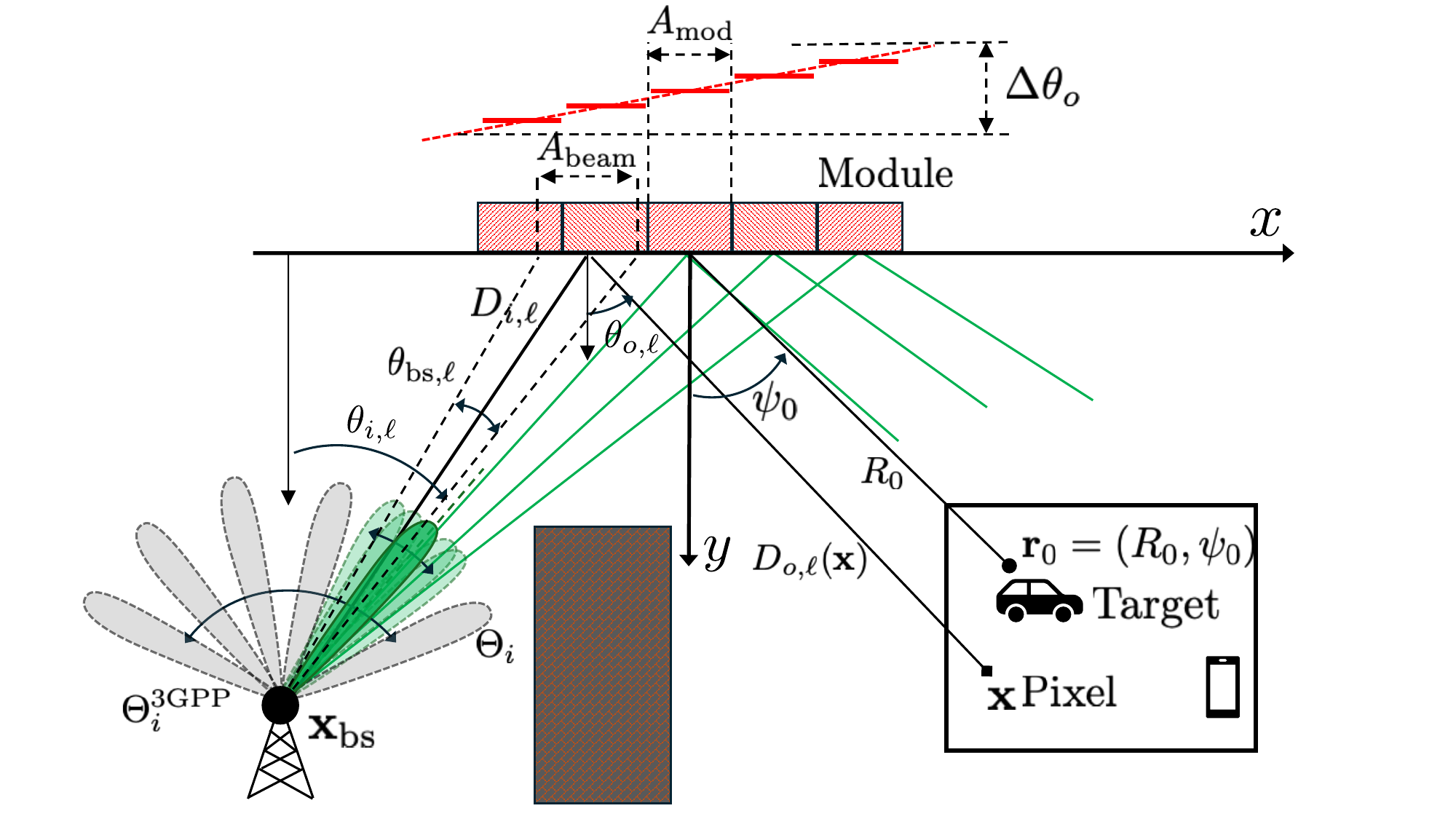}
    \caption{Sketch of the proposed system: the BS implements conventional IA (grey beams) with the addition of imaging-specific beams (green beams) to illuminate the reflector. The reflector is modular and allows exploring the ROI at high resolution. The coherent combination of return echoes on each beam gives the NLOS image. }
    \label{fig:scenario_BS}
\end{figure}
\section{System Design}\label{sec:system_design}

This section details the system design, from the selection of the BS beam codebook $\Theta_i$ to the design of the modular reflector and the derivation of the effective aperture. The geometry is illustrated in Fig. \ref{fig:scenario_BS}.

\subsection{Selection of the Tx Codebook $\Theta_i$ and IA Duration}
\label{ssec:tx_codebook}
The correct image of the targets in the ROI, obtained with  \eqref{eq:BP}, is achieved if two concurrent conditions hold: \textit{(i)} the Tx beam codebook $\Theta_i$ is designed to illuminate the reflector over multiple points (as explained in Section \ref{sec:imaging_principle}) and 
\textit{(ii)} the angular sampling within $\Theta_i$, here denoted with $\delta\theta_i$, is such that to ensure that grating lobes of the SAF $\chi[\mathbf{x}]$ fall outside the ROI. The Tx codebook $\Theta_i$ is:
\begin{equation}\label{eq:incidence_codebook}
    \Theta_i = \left\{\overline{\theta}_i -\frac{\Delta \theta_{i}}{2} : \delta \theta_i : \overline{\theta}_i + \frac{\Delta \theta_{i}}{2}\right\}
\end{equation}
where $\Delta \theta_{i}$ is such that to illuminate the entire reflector.
To prevent spatial aliasing, we analyze the variation of the instantaneous spatial frequency---defined as the rate of change of the propagation phase $\varphi (\mathbf{x})$ w.r.t. $\theta_i$ for a given pixel $\mathbf{x}$---with respect to an infinitesimal variation of the Tx angle $\theta_i$, obtaining: 
\begin{equation}\label{eq:ph_derivative}
\begin{split}
      \frac{\mathrm{d}\varphi (\mathbf{x})}{\mathrm{d}\theta_i } 
      \hspace{-0.1cm} = \hspace{-0.1cm}\frac{4 \pi D_y}{\lambda_0 \cos^2\theta_i} \hspace{-0.1cm}\left[ \sin\theta_i \hspace{-0.1cm}-\hspace{-0.1cm}\frac{x\hspace{-0.1cm}+\hspace{-0.1cm}D_x \hspace{-0.1cm}- \hspace{-0.1cm}D_y \hspace{-0.05cm}\tan \theta_i}{\sqrt{y^2 \hspace{-0.1cm}+\hspace{-0.1cm} (x\hspace{-0.1cm} + \hspace{-0.1cm}D_x \hspace{-0.1cm}-\hspace{-0.1cm} D_y \hspace{-0.05cm}\tan \theta_i)^2}}\right]
\end{split}
\end{equation}
where we have expressed the distances as function of $\theta_i$ and $\mathbf{x}$. It is a monotonically increasing function with $\theta_i$, as the path length increases with $\theta_i$ as well. The sampling interval is therefore ruled by the maximum difference between the phase derivatives within $\Delta \theta_{i}$, namely:
\begin{equation}\label{eq:deltathetai}
    \delta \theta_i \leq \frac{\pi}{\bigg \lvert \underset{\mathbf{x}}{\max} \left\{ \frac{\mathrm{d}\varphi (\mathbf{x})}{\mathrm{d}\theta_i } \big \rvert_{\frac{\Delta \theta_{i}}{2}}\right\} - \underset{\mathbf{x}}{\min} \left\{  \frac{\mathrm{d}\varphi (\mathbf{x})}{\mathrm{d}\theta_i } \big \rvert_{-\frac{\Delta \theta_{i}}{2}}\right\} \bigg \rvert}.
\end{equation}
In principle, the angular sampling $\delta \theta_i$ should guarantee that the corresponding sampling over the reflector is $\lambda_0/4$, as for conventional monostatic radar systems. However, considering a comparably small ROI relaxes the sampling condition. It is worth noticing that by doubling the ROI size or the reflector size ($A$, thus $\Delta \theta_i$), the angular sampling decreases to less than half of its value (a more than linear decrease), thus the required number of imaging beams $L\simeq \Delta \theta_i / \delta \theta_i$ grows more than linearly with $A$.  
The final Tx codebook employed by the BS for both IA and NLOS imaging is the union of the standard-compliant codebook $\Theta_i^{\rm 3GPP}$ and \eqref{eq:incidence_codebook}, as: 
\begin{equation}
    \Theta^{\rm IA}_i = \Theta_i \bigcup \Theta^{\rm 3GPP}_i.
\end{equation}
In this sense, the additional cost brought by the introduction of the NLOS imaging functionality into the IA is represented by the duration of the new IA procedure w.r.t the 3GPP compliant one:
\begin{align}
    T_{\rm IA, 3GPP}& =|\Theta^{\rm 3GPP}_i|T = K T \label{eq:DFT_IA}\\
    T_{\rm IA}&=|\Theta^{\rm IA}_i| T \simeq [L 
   + (K-\overline{K})] T \label{eq:proposed_IA}
\end{align}
where the first is function of the number of antennas $K$. The proposed IA for NLOS imaging adds the duration of the sweeping over the reflector (the $K-\overline{K}$ factor assumes that the orthogonal codebook illuminates the reflector with at most $\overline{K} = K \Delta \theta_i / (2 \pi /3)$ beams).

\subsection{Reflector Design and Effective Aperture } \label{ssec:ref_design}

The design of the reflector revolves around the selection of a proper set of reflection angles, $\Theta_o$, as introduced in Section \ref{sec:imaging_principle}. Some works design the phases of the meta-atoms by solving an optimization problem with a given cost function quantifying the difference between the desired and achieved electromagnetic field in the ROI~\cite{10621891}. However, direct optimization over large reflectors with thousands of meta-atoms is challenging and time-consuming~\cite{10926852}, thus it is generally advantageous in those cases where complex field patterns need to be implemented in the ROI.  
We propose a simple yet effective \textit{modular} approach, in which multiple subset of meta-atoms are independently phase-configured according to the well-known generalized Snell's law to implement a single reflection angle $\theta_o$ given $\theta_i$ (Fig. \ref{fig:scenario_BS}). This implementation allows us to approximate a NF design with a module-by-module FF design, and to obtain useful closed-form expression for the effective area $A_{\rm eff}$ and related insights. Notably, as demonstrated by our previous work~\cite{10926852}, the suitable configuration of a modular reflector (either static or reconfigurable) is substantially equivalent in terms of overall performance to the optimization of the whole set of meta-atoms when the reflection pattern needs to be uniform over a region, but with orders-of-magnitude less complexity. Fig. \ref{fig:scenario_BS} illustrates the proposed modular configuration of the reflector and the system.

Recalling \eqref{eq:angle_periodic_design_varyingincidence}, we shall be designing the angular function $f(x)$ in order to linearly span a contiguous set of angles covering the ROI. By denoting with $\overline{\theta}_o$ the angle from the center of the reflector to the center of the ROI and by $\Delta\theta_o$ the required angular span to illuminate the ROI, we have:
\begin{equation}\label{eq:angle_periodic_design_quant}
\begin{split}
    \theta_o(x) = \overline{\theta}_o \hspace{-0.1cm}+ \hspace{-0.1cm}\frac{\Delta \theta_o}{A} x   \approx \overline{\theta}_o \hspace{-0.1cm}+\hspace{-0.1cm} \sum_{n= 1}^{N} \frac{\Delta \theta_o}{A} \mathrm{rect}\left(\frac{x-x_n}{A_{\rm mod}}
    \right)  x_n
\end{split}
\end{equation}
where we have approximated $\theta_o(x)$ with a discrete set of $N$ modules, each of size $A_{\rm mod}=A/N$ and centered in $[x_n,0]$. The space-varying angular function to be implemented is then $f(x) = \theta_o(x) - \theta_i(x)$, where both the incidence and reflection angles are discrete over $N$ samples.
The number of meta-atoms per module is $M_{\rm mod} \simeq A_{\rm mod} / d$ and the phase configuration follows standard generalized Snell's law~\cite{7109827}\footnote{The phase applied to the $m$-th meta-atom of the $n$-th module to implement the discrete angle set \eqref{eq:angle_periodic_design_quant} is approximated as:
\begin{equation}
    \phi_m(x_n) = \frac{2 \pi}{\lambda_0} md \left[\sin \theta_{i,n} - \sin \theta_o(x_n)\right].\nonumber
\end{equation}}
.

Fixing an angular span in reflection $\Delta \theta_{o}$, the number of modules $N$ depends on manufacturing constraints. In general, having few large modules implementing a single angle eases manufacturing, as a linear phase gradient needs to be implemented. However, too large modules negatively affect imaging performance. To gain insight on this, let us refer to the geometry in Fig. \ref{fig:scenario_BS} and fix the module size $A_{\rm mod}$. The angular reflection pattern of the $n$-th module towards a generic angle $\psi$ is
\begin{equation}
    \eta(\psi) \hspace{-0,1cm}= \hspace{-0,1cm}\mathrm{sinc}\left[\frac{\sin \psi \hspace{-0,1cm} -\hspace{-0,1cm} \sin\theta_{o,n}}{\rho^{\rm mod}_{\psi}}\right],\,\, \rho^{\rm mod}_{\psi,n} = \frac{\lambda_0}{2 A_{\rm mod} \cos \theta_{o,n}},
\end{equation}
where $\rho^{\rm mod}_{\psi,n}$ is the angular resolution of the $n$-th module, function of the pointing angle $\theta_{o,n}=\theta_o(x_n)$.
Now, the key consideration is that the effective aperture $A_{\rm eff}$ for a given target in the ROI, that effectively rules the image resolution, is dictated by the number of modules whose reflection patterns overlap in the direction of the target, contributing to the target sensing. For a target in polar coordinates $\mathbf{r}_0=(R_0, \psi_0)$, we need to consider the radiation patterns $\{\eta(\psi_n)\}$, $n=1,..,N$, where $\psi_n = \arctan((r_x-x_n)/r_y)$ is the pointing angle of the $n$-th module towards the target. The set of modules that form the effective aperture $A_{\rm eff}$ is defined as
\begin{equation}\label{eq:Ceff}
    \mathcal{N}_{\rm eff} = \big\{n \; \big \lvert\; |\psi_n - \theta_{o,n}| \leq \rho^{\rm mod}_{\psi,n}\big\}
\end{equation}
and the effective aperture is $A_{\rm eff} = |\mathcal{N}_{\rm eff}| A_{\rm mod}$. 
We can derive the closed-form expression of the continuous set of points on the reflector constituting the effective aperture. By plugging in \eqref{eq:Ceff} the expression of the configuration angle $\theta_{o}(x)$ in \eqref{eq:angle_periodic_design_quant} and considering a generic NF condition, where we can expand the pointing angle as $\psi(x) \simeq \psi_0 - (\cos\psi_0/R_0) x$, we obtain the following continuous set of points on the reflector:
\begin{equation}\label{eq:xeff}
    \mathcal{X}_{\mathrm{eff}} =
\left\{ x \in 
\frac{  (\psi_0-\bar{\theta}_o) \mp \dfrac{\lambda_0}{2A_{\mathrm{mod}}\cos\bar{\theta}_o}}
     {\dfrac{\Delta\theta_o}{A}\!\left[1\pm\dfrac{\lambda_0\tan\bar{\theta}_o}{2A_{\mathrm{mod}}\cos\bar{\theta}_o}\right]
     + \dfrac{\cos\psi_0}{R_0}}
\right\},
\end{equation}
forming a segment whose inferior endpoint is achieved with the minus sign at the numerator and the plus sign at the denominator, and vice-versa for the superior endpoint. 
The effective aperture $A_{\rm eff}$ is then the length of the segment identified by \eqref{eq:xeff}, limited to the reflector size $A$ ($A_{\rm eff}\leq A$). This has the following interpretations.


First, for fixed ROI position w.r.t. the reflector (fixed $\cos\psi_0/R_0$), increasing the ROI size means increasing the angular span $\Delta \theta_o$, and the effective aperture $A_{\rm eff}$ linearly decreases. This strikes an intuitive and explicit trade-off: the larger the area to be imaged, the less is the effective resolution.  

Second, for fixed ROI size and position w.r.t. the reflector (fixed $\cos\psi_0/R_0$ and $\Delta\theta_o$), increasing the reflector size $A$ (let $A\rightarrow \infty$) increases the effective aperture up to a limit due to $\cos\psi_0/R_0$. This means that only a finite number of modules contribute to the image resolution, and the ones located at the off-boresight w.r.t. target are useless, as expected from Huygens diffraction principle.

Third, by increasing the \textit{electromagnetic size} of the modules $A_{\rm mod}/\lambda_0$, narrowing its reflection beam, the effective area decreases. This is demonstrated by computing the derivative of the effective area $A_{\rm eff}$ w.r.t. the module size $A_{\rm mod}$\footnote{The derivative is proportional to  
    \begin{equation}
        \frac{\partial A_{\rm eff}}{\partial A_{\rm mod}} \propto - \left(\frac{\Delta \theta_o}{A} \left[1- (\psi_0-\overline{\theta}_o) \tan \overline{\theta}_o\right] + \frac{\cos \psi_0}{R_0}\right)\nonumber
    \end{equation}
    and it is negative when $(\psi_0-\overline{\theta}_o) \tan \overline{\theta}_o < 1$, that holds in typical conditions (comparably small ROI, not located at off-boresight where the resolution gets to infinite).}. In the limiting case, there is only one module that illuminates the target in $\mathbf{r}_0$, i.e., $|\mathcal{N}_{\rm eff}|\rightarrow 1$. In such scenario, the effective aperture reduces to:
    \begin{equation}
    A_{\rm eff} = \begin{dcases} A_{\rm bs} & \text{if}\, A_{\rm beam} \leq A_{\rm mod}   \\
    A_{\rm mod} & \text{if}\, A_{\rm beam} > A_{\rm mod}
    \end{dcases}
    \end{equation}
    depending on the beam projection area $A_{\rm beam}$.
    In the first condition ($A_{\rm beam} \leq A_{\rm mod}$), the system is equivalent to an anomalous mirror, and the resolution is dictated by the BS aperture. In the second condition ($A_{\rm beam} > A_{\rm mod}$), only a portion of the BS beam is reflected, and the image resolution is ruled by the single module size. 

Conversely, by decreasing the module size $A_{\rm mod}/\lambda_0$, thus broadening its reflection beam such that $|\mathcal{N}_{\rm eff}| > 1$, the effective aperture tends to \textit{increase}. 
This latter phenomenon is counterintuitive, but can be explained by noticing the parallelism between the proposed NLOS imaging system based on beam sweeping and a synthetic aperture radar system. For such systems, the smaller is the physical antenna (module) the larger is the synthetic (effective) aperture, as the objective is to sense the environment from the wider possible angular interval. Remarkably, using reasonably small modules allows attaining the \textit{lens limit}, i.e., $A_{\rm eff} \rightarrow A$, over moderate-size ROIs. By using \eqref{eq:xeff} in \eqref{eq:reoslution_NF}, we obtain a closed-form expression for the image resolution in NF, not reported for brevity. 
As a last comment, we want to highlight that this reflector configuration is optimal for imaging in the sense that the effective aperture is composed of \textit{contiguous} modules, thus the final coherent image is not affected by high sidelobes (we avoid sparse array effects). Other non-linear angular configurations $f(x)$ are possible, though they yield effective apertures that are undesirably target-dependent.

\section{Imaging of Moving Targets}
\label{sec:moving_target_imaging}



Recalling Eq. \eqref{eq:BP}, forming the correct image of a moving target requires the prior knowledge of its velocity, in order to compensate for the propagation phase $\varphi_\ell (\mathbf{x},\mathbf{v})$ and enable the coherent---constructive---sum of received echoes at each beam. If the velocity of the target is not considered, the test phase \eqref{eq:test_phase} does not match the true propagation carrier phase and the final image suffers from quality degradation, detailed in the following.

    

\vspace{-0.2cm}\subsection{Image Degradation due to Target's Motion}\label{eq:image_degradation}

To quantify the detrimental effects of the target's motion on the final image, let us consider a single target in the ROI, located in $\mathbf{r}_0 = (R_0,\psi_0)$ and moving with velocity $\mathbf{v}_0=(v_R, v_T)$, where the latter is parameterized as radial (along $\psi_0$) and transverse (across $\psi_0$). Since the target can be in the NF of the whole reflector, the spherical wavefront over the reflector can be approximated as parabolic  w.r.t. $x_\ell$: 
\begin{equation}\label{eq:Do_NF}
    D_{o,\ell} = \|\mathbf{r}_0 - \mathbf{p}_\ell \| \simeq R_0 - \sin \psi_0 x_\ell  + \frac{\cos^2\psi_0}{2 R_0} x^2_\ell .
\end{equation}
Likewise, we expand the Doppler $\nu_\ell$ at the first order w.r.t. $x_\ell$---i.e, observing the first order variations of the Doppler shift across the sweeping over the reflector---, obtaining:
\begin{equation}\label{eq:Doppler_first_order}
\begin{split}
    \nu_\ell \hspace{-0.1cm} = \hspace{-0.1cm} -\frac{2}{\lambda_0} \frac{\mathbf{v}_0^T \hspace{-0.05cm}\left(\mathbf{r}_0\hspace{-0.1cm} -\hspace{-0.1cm} \mathbf{p}_\ell\right)}{\|\mathbf{r}_0\hspace{-0.1cm} - \hspace{-0.1cm}\mathbf{p}_\ell\|}
    & \simeq  \hspace{-0.05cm} - \frac{2}{\lambda_0} \left[v_R \hspace{-0.1cm}+ \hspace{-0.1cm}\frac{v_T \cos \psi_0}{R_0} x_\ell \right].
\end{split}
\end{equation} 
The first term is proportional to the radial velocity, i.e., $v_R = \mathbf{v}^T \mathbf{u}_R$, with $\mathbf{u}_R=[\sin\psi_0,\cos\psi_0]^T$, and independent on $\ell T$ (yielding a linear phase). The second term depends on the transverse velocity $v_T = \mathbf{v}^T \mathbf{u}_T$, with $\mathbf{u}_T=[-\cos \psi_0 , \sin\psi_0]^T$ and it is linear in $x_\ell$. By approximating $x_\ell$  as $x_\ell = v_{\rm sweep} \ell \Delta t$, i.e., assuming a constant sweep velocity over the reflector, we have a quadratic phase trend in $\ell$. The sweeping velocity $v_{\rm sweep}$ is 
\begin{equation}
    v_{\rm sweep} = \frac{1}{L}\sum_{\ell}\frac{x_{\ell+1}-x_\ell}{T}
\end{equation}
and it represents the average beam sweeping speed over the reflector. Notice that in FF we have a single term in the Doppler expression, depending only on the radial velocity $v_R$. The transverse velocity $v_T$ is not observable in FF due to the lack of wavefront curvature across the reflector. 

If there is no \textit{range migration} of the target during the observation time, i.e., $\|\mathbf{v}\| L T \ll \rho^{\rm (NF)}_R, R_0 \rho^{\rm (NF)}_\psi$, the (noiseless) image in polar coordinates is reported in \eqref{eq:image_polar_coordinates_NF},
\begin{figure*}
\begin{equation}\label{eq:image_polar_coordinates_NF}
    \begin{split}
        I(R,\psi) &\overset{\rm(NF)}{\approx}  \alpha_0\, \mathrm{sinc}\left[\frac{R \hspace{-0.1cm}-\hspace{-0.1cm}R_0}{\rho_R}\right]
        \sum_\ell 
        e^{j \frac{4 \pi}{\lambda_0} \left( \left(v_{\rm sweep}[\sin\psi_0-\sin\psi] - v_R\right)  \ell T + \left(v^2_{\rm sweep} \left[\frac{\cos^2\psi_0}{2R_0} - \frac{\cos^2\psi}{2R}\right] - \frac{v_T}{R_0}\cos\psi_0 v_{\rm sweep}\right) \ell^2 T^2  \right)} \\
        &\overset{\rm(FF)}{\approx} \alpha_0\,
        \mathrm{sinc}\left[\frac{R \hspace{-0.1cm}-\hspace{-0.1cm}R_0}{\rho_R}\right]\mathrm{sinc}\left[\frac{ 2 \pi L_{\rm eff} T\left(v_{\rm sweep} \left[\sin\psi_0 \hspace{-0.1cm}-\hspace{-0.1cm}\sin\psi \right] \hspace{-0.1cm}- \hspace{-0.1cm}v_R \right)}{\lambda_0}\right]
    \end{split}
\end{equation}
    \hrulefill
\end{figure*}
where the first line is the NF expression and the second is the FF approximation. In both cases, the summation over subcarriers is not affected by Doppler, and it yields a sinc function along $R$. In FF, the image has a closed form expression, and the peak of the cardinal sine along azimuth is the \textit{apparent} angular position \cite{manzoni2022motion}:
\begin{equation}
    \widehat{\psi}_0 = \arcsin{\left( \sin\psi_0 - \frac{v_R}{v_{\rm sweep}} \right)}.
\end{equation}
A non-compensated radial velocity of the target $v_R$ gives a rotated image by $|\widehat{\psi}_0-\psi_0|$. Since each target may have different velocity, the image is distorted. The amount of angular rotation depends on $v_R/v_{\rm sweep}$, thus: \textit{(i)} the longer the sweep over the reflector, the more is the impact of target motion on the final image, and it cannot be neglected; \textit{(ii)} if $v_{\rm sweep} \gg v_R$, the rotation is negligible, but the accuracy on velocity estimation lowers as well. If the BS aims at estimating the targets' velocities, the sweeping shall have a minimum duration. 

In NF, instead, there is not a closed form solution for the image. We have the same image rotation due to $v_R$, and an additional quadratic term, that is due to the curvature of the wavefront and the transverse velocity $v_T$. However, a quadratic phase can be approximated with a finite set of linear phases pertaining to as many sub-apertures (e.g., of the module size), that yield progressive image rotations. Therefore, the final sum of slightly rotated and phase-shifted images gives \textit{defocusing} (blurring), i.e., a degradation of the SAF with amplitude and resolution reduction.

\vspace{-0.2cm}\subsection{Imaging of Moving Targets}

For targets in FF, only their radial component of the velocity is needed, while NF requires the prior knowledge of the 2D velocity vector of the targets. Therefore, we need to estimate the targets' velocity vector \textit{before} image formation \eqref{eq:BP}. The general approach in NF is by ML methods, herein adopted, starting from the samples of the residual propagation phase on the single-beam low-resolution---often referred to \textit{pre-stack}, pre-summation---images. Single-beam images are defined as
\begin{equation}
I_\ell(\mathbf{x}) \hspace{-0.1cm}= \hspace{-0.1cm}\sum_q Y_{q,\ell} \frac{S_{q,\ell}^* \beta^*_\ell(\mathbf{x}) G^*_\ell(\mathbf{x})}{\sqrt{P_{\rm tx}}|\beta_\ell(\mathbf{x})|^2 |G_\ell(\mathbf{x})|^2} e^{j 2 \pi (f_0+q \Delta f) \tau_\ell(\mathbf{x})}
\end{equation}
whose resolution is roughly dictated by the bandwidth and either the BS array size or the module size (Section \ref{sec:system_design}). Let us assume that the $U$ targets are detectable in the single images. Their position can be coarsely estimated using conventional approaches, this gives $U$ estimated target's positions in each of the $L$ images, that can be averaged to obtain $\widehat{\mathbf{r}}_u = \sum_\ell \widehat{\mathbf{r}}_{u,\ell}/L = (\widehat{R}_u,\widehat{\psi}_u)$, $u=1,...,U$. The estimated position of the target is assumed to be sufficiently accurate for velocity estimation.
By designing the reflector as in Section \ref{sec:system_design}, we can sample the phase of the $L$ single-beam images in $\widehat{\mathbf{r}}_u$, and unwrap them obtaining the vector $\widehat{\phi}_{u,\ell} = \angle I_\ell(\widehat{\mathbf{r}}_u)$, $\ell=-L/2,..,L/2-1$. The unwrapped phase vector $\widehat{\boldsymbol{\phi}}_u = [\widehat{\phi}_{u,\ell}] \in \mathbb{R}^{L\times 1}$ is a function of the velocity $\mathbf{v}_u$ of the $u$-th target as follows:
\begin{equation}\label{eq:phase_model_quadratic}
     \widehat{\boldsymbol{\phi}}_u = \Phi_u \hspace{-0.05cm}+\hspace{-0.05cm} \underbrace{\frac{4\pi}{\lambda_0} v_{R,u} T}_{a_1} \boldsymbol{\ell} + \underbrace{\frac{4\pi}{\lambda_0} \frac{ v_{T,u} \cos \widehat{\psi}_{u}}{\widehat{R}_{u}} v_{\rm sweep}T^2}_{a_2} \boldsymbol{\ell}^2  + \mathbf{n}_\phi
\end{equation}
where we defined $\boldsymbol{\ell} = [-L/2,..,L/2-1]^T$ the vector of beam indices and $\boldsymbol{\ell}^2 = \boldsymbol{\ell}\odot \boldsymbol{\ell}$. The noise on phase measurements is $\mathbf{n}_\phi \sim \mathcal{N}(\mathbf{0}, \mathbf{C}_\phi)$, with covariance matrix $[\mathbf{C}_\phi]_\ell = 1/(2 \mathsf{SNR}_{u,\ell})$. The coefficients $a_1$ and $a_2$ are estimated with the ML approach as $\widehat{\mathbf{a}} = (\mathbf{L}^T\mathbf{C}_\phi^{-1} \mathbf{L})^{-1} \mathbf{L}^T \mathbf{C}_\phi^{-1}\widehat{\boldsymbol{\phi}}_u$, where we define the regression matrix $\mathbf{L} = [\boldsymbol{\ell}, \boldsymbol{\ell}^2]$. 
The CRB has the well-known expression for linear systems in Gaussian noise, i.e., $\mathbf{C}_{\mathbf{a}} = (\mathbf{L}^T \mathbf{C}_\phi^{-1}\mathbf{L})^{-1}$, and, in the proposed system, it can be approximated by considering the subset $\mathcal{L}_{\rm eff} \subseteq \{-L/2,...,L/2-1\}$ of effective beams. As per polynomial regression, the variance of $a_1$ scales with $L_{\rm eff}^{-3}$ (frequency estimation), while the variance of $a_2 \sim L_{\rm eff}^{-5}$.  
The resulting velocity components are then:
\begin{equation}
    \widehat{v}_{R,u} = \frac{\lambda_0}{4 \pi T} \widehat{a}_1\,\,\,\, \widehat{v}_{T,u} = \frac{\lambda_0 \widehat{R}_{u}}{4 \pi T (v_{\rm sweep}T) \cos \widehat{\psi}_{u} }\widehat{a}_2.
\end{equation}
The estimated velocities are plugged in \eqref{eq:Doppler_first_order} to generate the Doppler-corrected images. 
The accuracy on transverse velocity $v_{T,u}$ decreases with increasing radial distance $R_u$. This is expected since increasing $R_u$ for fixed $A$ diminishes the angular diversity in the observation of the target's motion. This result is consistent with the literature \cite{Giovannetti_NF_velocity}. Notice that the estimated transverse velocity depends on the rough initial estimate of the target's position $\widehat{\mathbf{r}}_u = (\widehat{R}_u,\widehat{\psi}_u)$ and on the resulting bias $\boldsymbol{\epsilon}_u = \widehat{\mathbf{r}}_u - \mathbf{r}_u$, inducing high-order terms in \eqref{eq:phase_model_quadratic} that are here omitted for simplicity. The bias $\boldsymbol{\epsilon}_u$ depends on the resolution of the single images, that might be very low depending on specific settings. The proposed procedure can be however iterated from the Doppler-corrected images, obtained using $\widehat{v}_{R,u}$ and $\widehat{v}_{T,u}$, repeating the estimate $\widehat{\mathbf{r}}_u$ and so on. If the BS uniformly samples the reflector with constant velocity (as in \eqref{eq:phase_model_quadratic}), other low-cost approaches are feasible, such as \cite{FrFT}. 
The proposed image-based velocity estimation approach is also useful to compensate for unknown orientations of the reflector, whose effect is similar to a non-compensated radial velocity.

\section{Three Trade-offs and One Challenge for Integrating Imaging in 6G IA}
\label{sec:tradeoffs}

In practice, the integration of NLOS imaging into the 6G IA procedure brings new challenges and trade-offs.
The main challenge is the limited bandwidth of the SSB signal. Standard-compliant (5G) SSB used for IA employs up to 240 continuous subcarriers, thus a useful bandwidth ranging from $3.6$ MHz ($\mu=0$, $\Delta f=15$ kHz) to $57$ MHz ($\mu=4$, $\Delta f=240$ kHz)~\cite{Giordani_beammanagement}, yielding a range resolution of $\rho_R\simeq 41$ m to $\rho_R\simeq 2.6.$ m. The latter values, limited by the SSB bandwidth, are not sufficient for high-resolution imaging purposes. Among possible solutions, we can consider: \textit{(i)} exploiting the remaining unused subcarriers to opportunistically transmit user data, as the 3GPP standard does not mention any limitation on this. This approach is however limited by the effective availability and DL data, and the resulting bandwidth varies from beam to beam (SSB to SSB); \textit{(ii)} filling the unused subcarriers with dedicated pilots, that are not necessarily known at the user side; \textit{(iii)} use a reflector large enough to compensate for the lack of bandwidth with the NF effect. Regarding the latter, however, Eq. \eqref{eq:range_factor_NF} shows that for fixed fractional bandwidth $B/f_0$ the required ratio $A/R_0$ should be overwhelming. For example, at $f_0=10$ GHz, $B=30$ MHz (SSB with $\Delta f =128$ kHz), in order to have a bandwidth enhancement factor $\kappa_R=10$ for a target at $R_0=10$ m, $\psi_0=0$, we need a reflector of effective aperture of $A_{\rm eff} \simeq 5 $ meters. This has a dramatic impact on IA duration, since increasing $A$ (and in turn $A_{\rm eff}$) leads to a consequent increase of the number of required additional beams $L$. Here, we consider option \textit{(ii)}, filling unused portions of the spectrum with pilot subcarriers, only on the $L$ imaging beams, and use a reflector of moderate size. In general, a linear increase in the reflector size $A$ maps into a more than linear increase in $T_{\rm IA}$. A similar consideration can be made for the SNR, as per \eqref{eq:SNR_coherent}, an increase in SNR in the final image calls for using narrower beams (limited by BS array size), longer pilot signals (limited by standard regulations) or denser Tx codebooks $\Theta_i$ across the reflector. Again, this increases the IA duration $T_{\rm IA}$. A linear increase in $L$ maps into a linear increase in the SNR. Lastly, increasing the accuracy on velocity estimation implies using either a denser codebook $\Theta_i$ or a larger reflector, to increase the observation time through number of samples $L$, affecting the IA duration. Therefore, the design of the system is a balanced trade-off between imaging and communication performance, that can be dependent on the context.

\section{Results} 
\label{sec:results}



This section details the performance evaluation of our system, as we performed both numerical simulations and experimental field tests in a controlled indoor environment. For numerical simulations, we considered a BS equipped with an ULA of size $A_{\rm bs}=0.4$ m operating at $f_0 \in \{15,28\}$ GHz as for frequency range 3 (FR3) and FR2 bands. The bandwidth is $B = 400$ MHz at $f_0=28$ GHz and $B = 200$ MHz at $f_0=15$ GHz. The geometry is such that the BS illuminates the reflector (whose size $A$ is let vary in simulations) by Tx angle $\overline{\theta}_i = 20$ deg \, at distance $D_y=5$ m. The target ROI is located at distance $R_0 = 15$ m and angle $\psi_0 = 0$ deg from the center of the reflector. The Tx power per antenna is $P_{tx} = 1$ W. The signal at the Rx side is corrupted by thermal noise with power $\sigma_w^2 = N_0B$, $N_0 = -173$ dBm/Hz. We fix the 3GPP numerology to have $T=0.25$ ms, $\Delta t=71.5$ $\mu$s.

\begin{figure}
    \centering
    \subfloat[][]{\includegraphics[width=\linewidth]{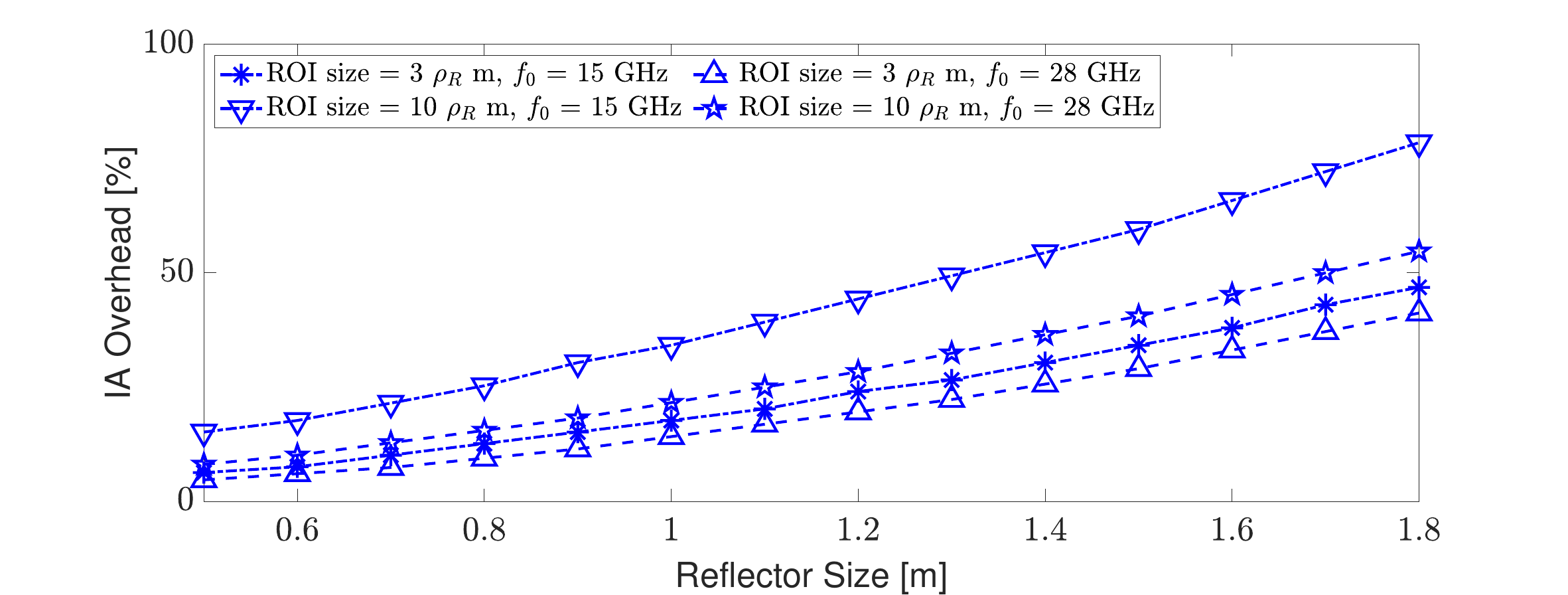}} \vspace{-0.3cm}\\
    \subfloat[][$f_0=15$ GHz]{\includegraphics[width=\linewidth]{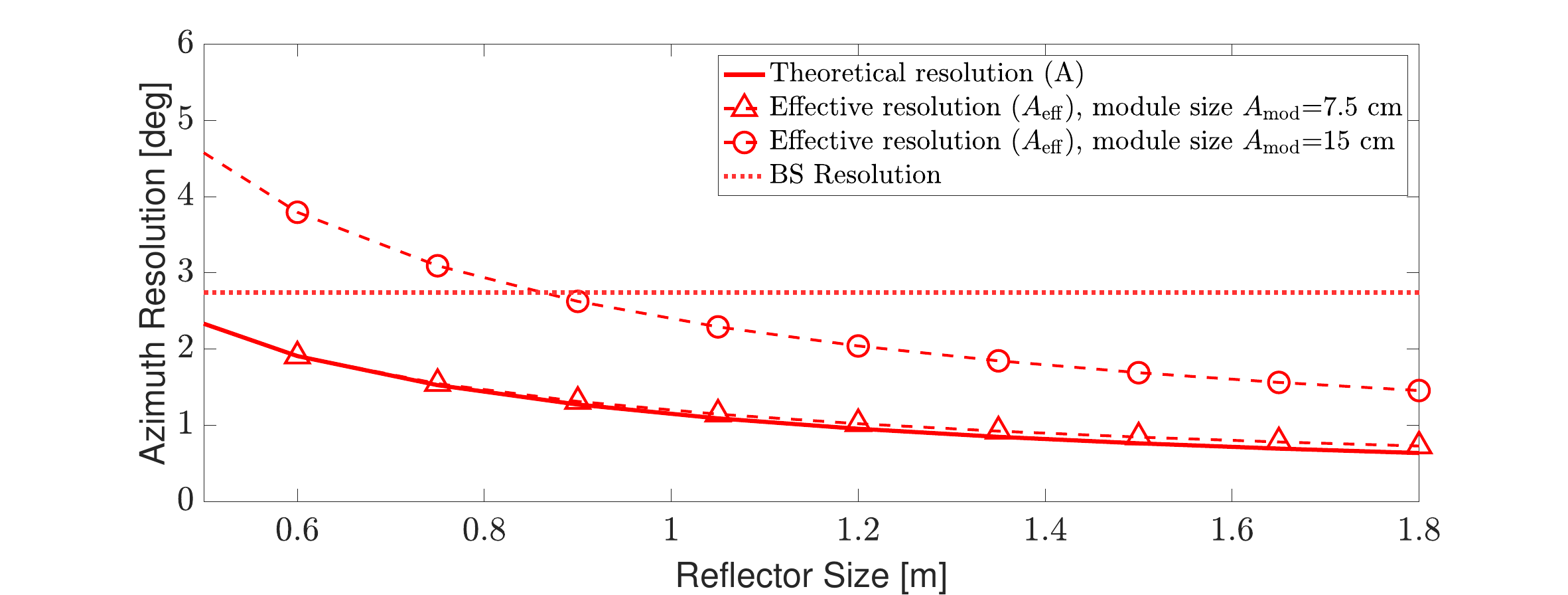}}\vspace{-0.3cm}\\
    \subfloat[][$f_0=28$ GHz]{\includegraphics[width=\linewidth]{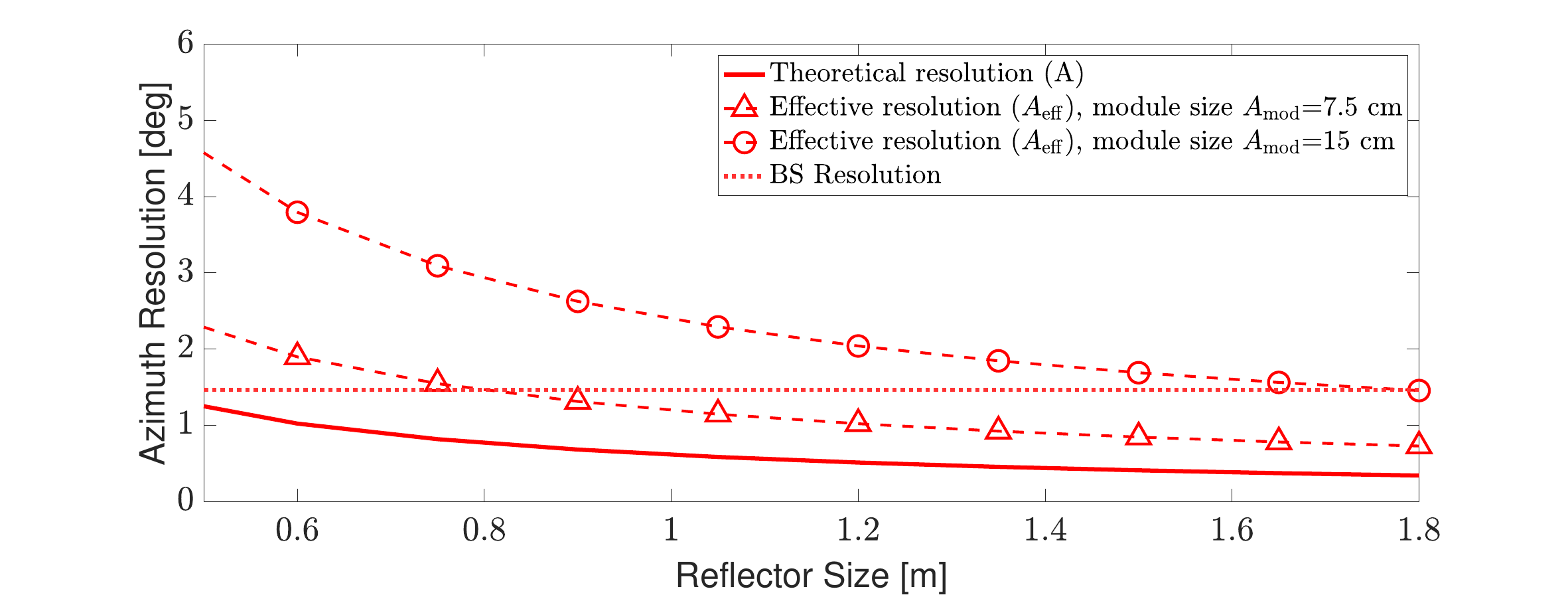}}\vspace{-0.3cm}\\
    \subfloat[][Radial velocity $v_R$]{\includegraphics[width=\linewidth]{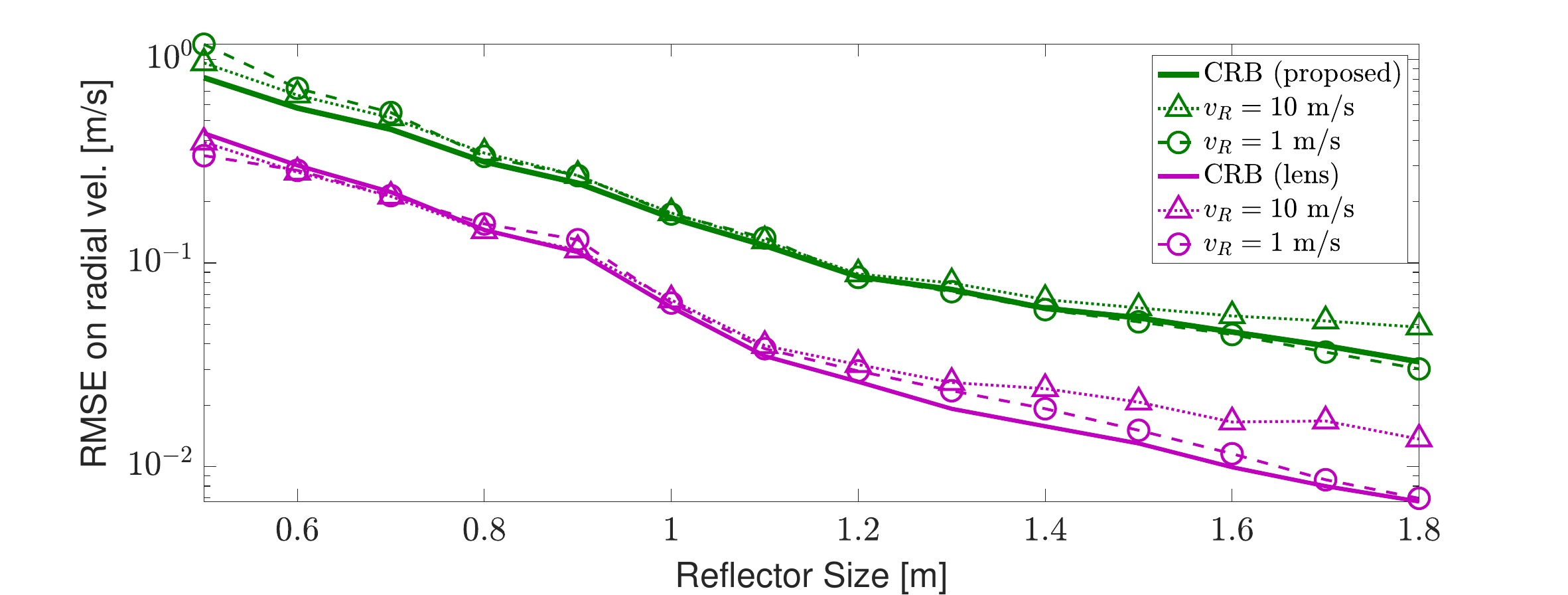}}\vspace{-0.3cm}\\
    \subfloat[][Transverse velocity $v_T$]{\includegraphics[width=\linewidth]{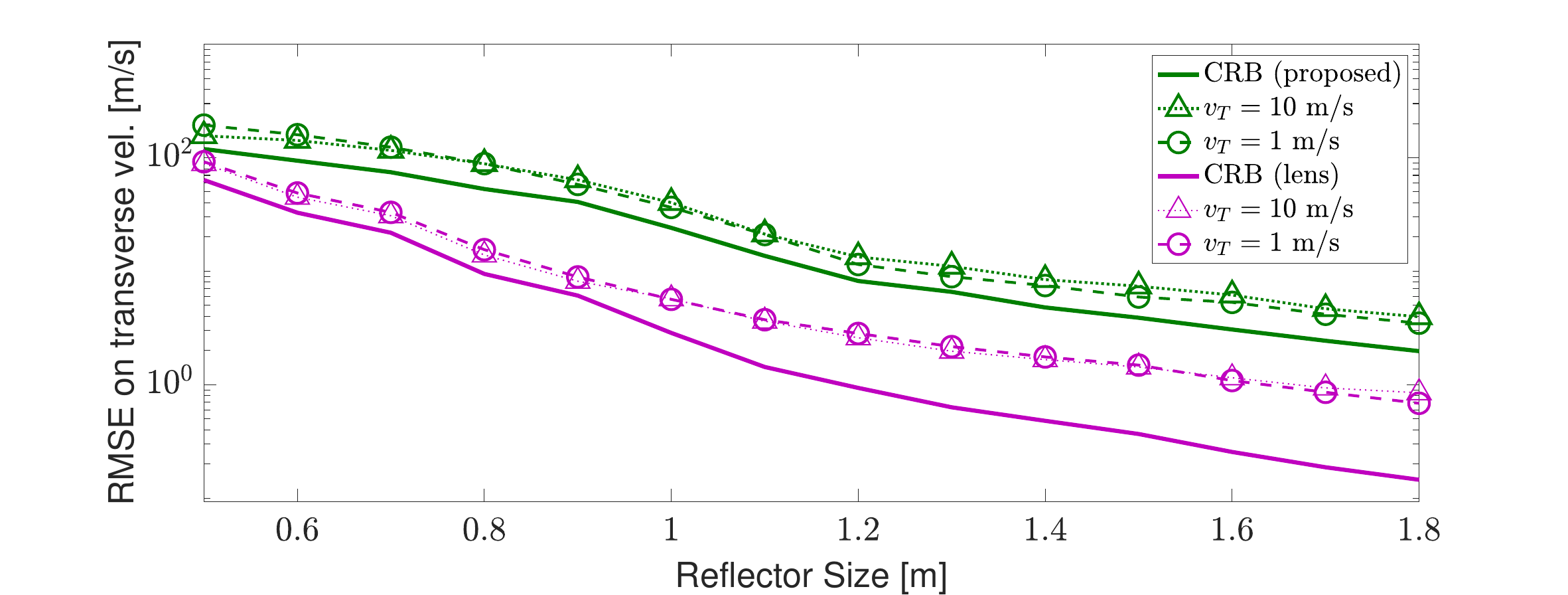}}
    \caption{Performance trade-off from the integration of NLOS imaging into IA: (a) IA duration overhead (b,c) Azimuth resolution (d,e) RMSE on velocity estimation.}
    \label{fig:trade-off}
\end{figure}

\subsection{Performance Trade-Offs}

The first result, in Fig. \ref{fig:trade-off}, reports the performance trade-off arising from introducing NLOS imaging into the 3GPP IA procedure. We study the trade-off against the reflector size $A$, as this is the main parameter driving the performance. Fig. \ref{fig:trade-off}a shows the IA overhead, expressed as the percentage increase with respect to the baseline 3GPP codebook duration, i.e. $\left(T_\text{IA}/T_\text{IA,3GPP}\right)\cdot 100$. We show the IA overhead for two ROI size, $\Delta_x = \Delta_y \in \{ 3\rho_R, 10 \rho_R\}$, where $\rho_R = c/(2B)$ is the FF range resolution achieved with bandwidth $B$, changing with $f_0$, to allow a fair comparison between the two systems. According to \eqref{eq:DFT_IA}-\eqref{eq:proposed_IA}, the proposed IA duration is ruled by the number of additional beams $L$ and by the number of BS antennas $K$, ruling the cardinality of the 3GPP codebook. The former increases more than linearly with both the reflector size $A$ and the ROI size, justifying the trend in Fig. \ref{fig:trade-off}a, as many beams are needed to avoid grating lobes within the ROI when increasing $A$. The number of BS antennas $K$, instead, scales linearly with the carrier frequency for fixed BS aperture $A_{\rm bs}$, and so does the 3GPP IA duration. Interestingly, at FR2 ($28$ GHz) we have a lower overhead compared to FR3 ($15$ GHz), since 3GPP IA already uses a denser codebook, and the required addition of $L$ beams has a lesser impact. Therefore, increasing the reflector size $A$ and the ROI size increases the overhead, but lower frequencies are more affected by NLOS imaging, up to $80\%$ increase in the IA duration in the considered settings. It is worth remarking that, in practice, the consequent capacity reduction due to IA extension depends on its periodicity; a detailed analysis is however not reported for brevity.

In practice, the IA overhead is justified by the beneficial enhancement of the angular resolution $\rho_\psi$. Fig. \ref{fig:trade-off}b and \ref{fig:trade-off}c shows the angular resolution as per \eqref{eq:reoslution_NF} using the expression of the effective area $A_{\mathrm{eff}}$ from \eqref{eq:xeff}, for the two carrier frequencies and the larger ROI. We show the achievable BS resolution (using effective aperture $A_{\rm bs}$, dotted line), the theoretical maximum resolution (when the reflector is configured as a lens toward the target position, solid line) and the effective resolution for two values of module size $A_{\rm mod}$ (marked lines).
As expected, increasing $A$ reduces $\rho_\psi$. The theoretical (lens) resolution is the lower bound, that is attained only at FR3 ($f_0=15$ GHz) for small modules, while FR2 requires even smaller modules, challenging the manufacturing of the reflector. The theoretical resolution is always better than the BS one in the considered settings, while the \textit{effective} one depends on module size and, if $A_{\rm mod}$ is too large, it attains the BS resolution only for very large reflectors (Fig. \ref{fig:trade-off}c).  
In general, the resolution enhancement is more pronounced at lower carrier frequencies. As a rule of thumb, comparing Fig. \ref{fig:trade-off}a and \ref{fig:trade-off}b shows that to halve the angular resolution $\rho_\psi$ compared to BS one, the IA overhead is $\approx 60\%$.

By increasing the reflector size we also enhance the accuracy on velocity estimation. Fig. \ref{fig:trade-off}d shows the root mean squared error (RMSE) on radial velocity $v_R$, and \ref{fig:trade-off}e the RMSE on the transverse one $v_T$. Solid lines show the CRB, while dashed and dotted lines the RMSE, for the reflector configured as proposed (green curves) or lens (purple curves, lower bound), at $f_0=15$ GHz. The SNR on the single phase measurement (over the effective samples) ranges from $18$ to $21$ dB in the considered setup. The proposed system estimates the radial velocity with cm/s accuracy with large reflectors, while transverse velocity, only observable through the wavefront curvature across the reflector, is estimated with higher error, roughly two orders of magnitude. The RMSE follows the trend of the CRB and the gap between the two increases with $A$, especially for the lens configuration. This is due to residual high-order non-linear terms not modeled in the phase \eqref{eq:phase_model_quadratic} used to estimate the velocity. These are due to the small sampling bias $\boldsymbol{\epsilon}=\widehat{\mathbf{r}}-\mathbf{r}$ and mild range migration effects, occurring for large velocities and apertures $A$. It is worth mentioning that, in the context of the paper, the estimated velocities are instrumental to correct the image distortions outlined in Section \ref{eq:image_degradation}. Residual biases due to high-order terms are however low enough to have a quantitative assessment of the true velocity of the target.

\begin{figure*}[t] 
    \centering
  \subfloat[\label{fig:sub1}]{
    \includegraphics[width=0.23\columnwidth]{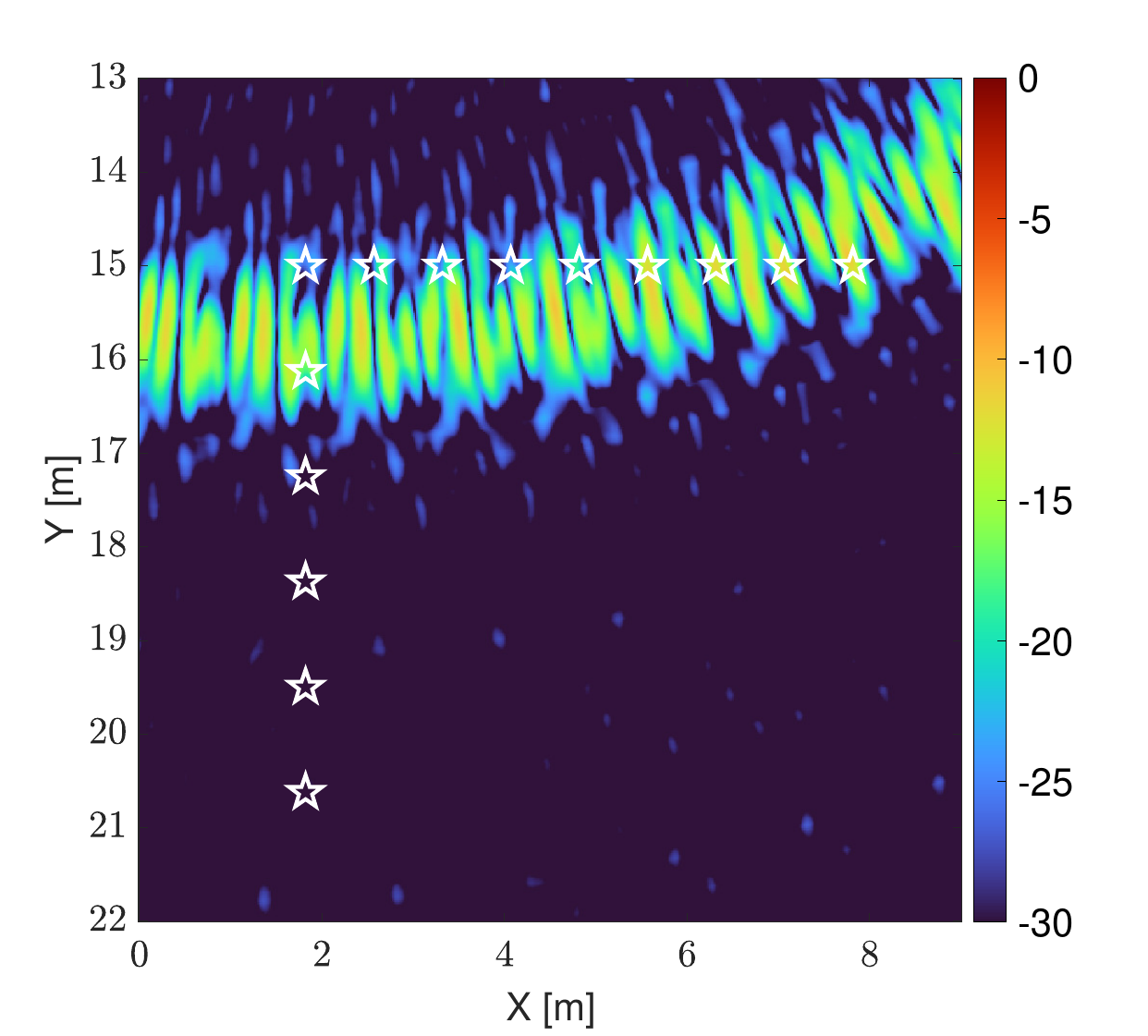}
  }
  \hfill
  \subfloat[\label{fig:sub2}]{
    \includegraphics[width=0.23\columnwidth]{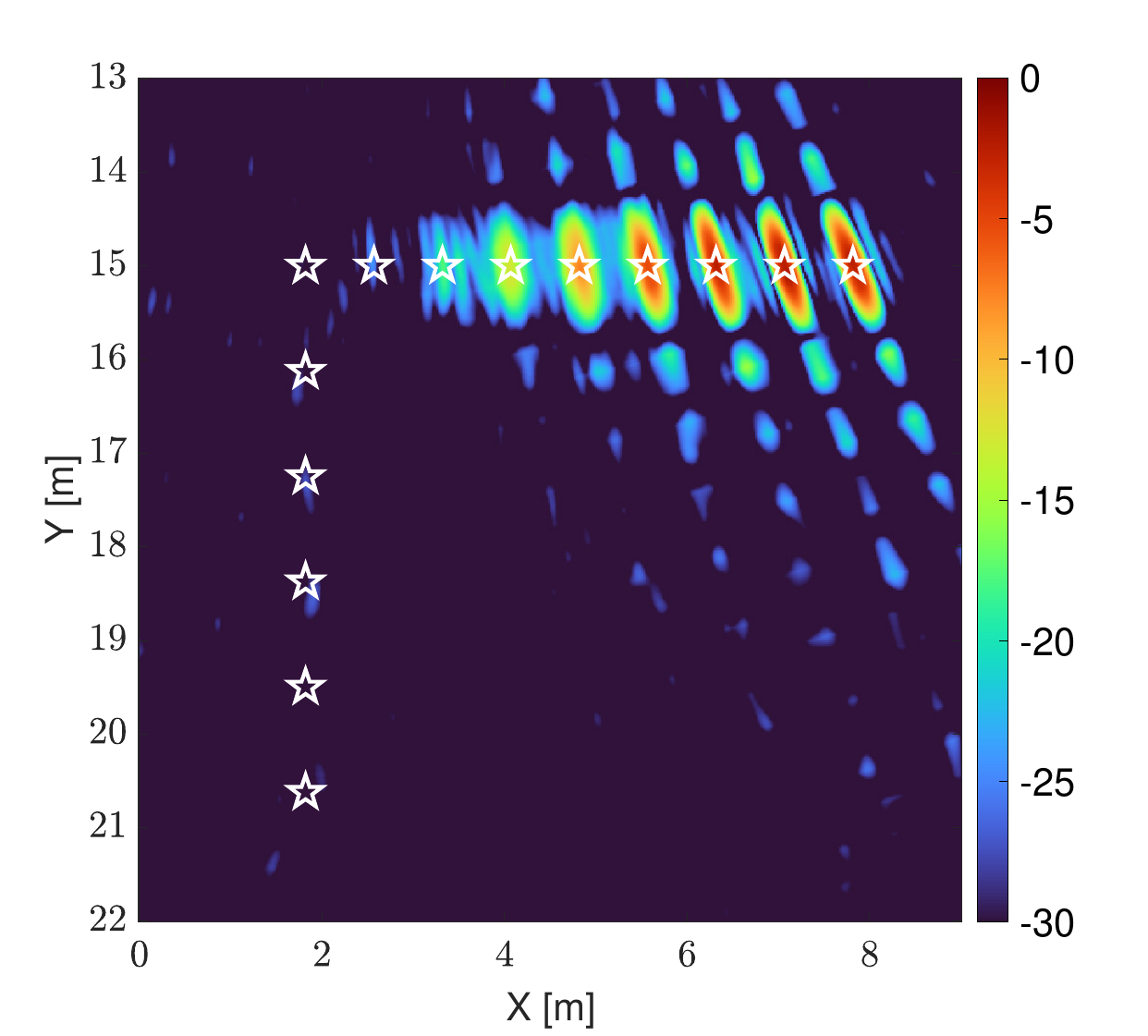}
  }
  \hfill
  \subfloat[\label{fig:sub3}]{
    \includegraphics[width=0.23\columnwidth]{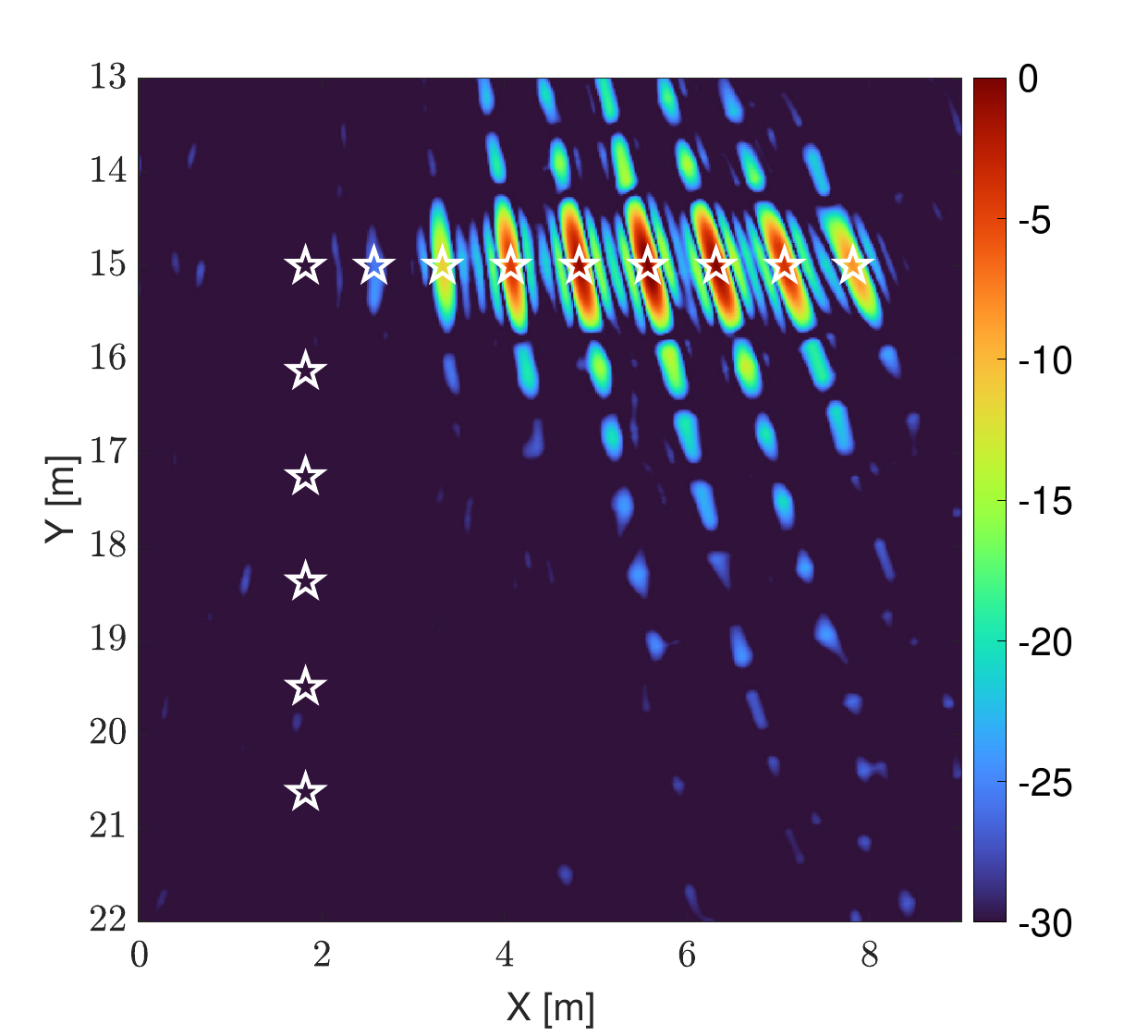}
  }
  \hfill
  \subfloat[\label{fig:sub4}]{
    \includegraphics[width=0.23\columnwidth]{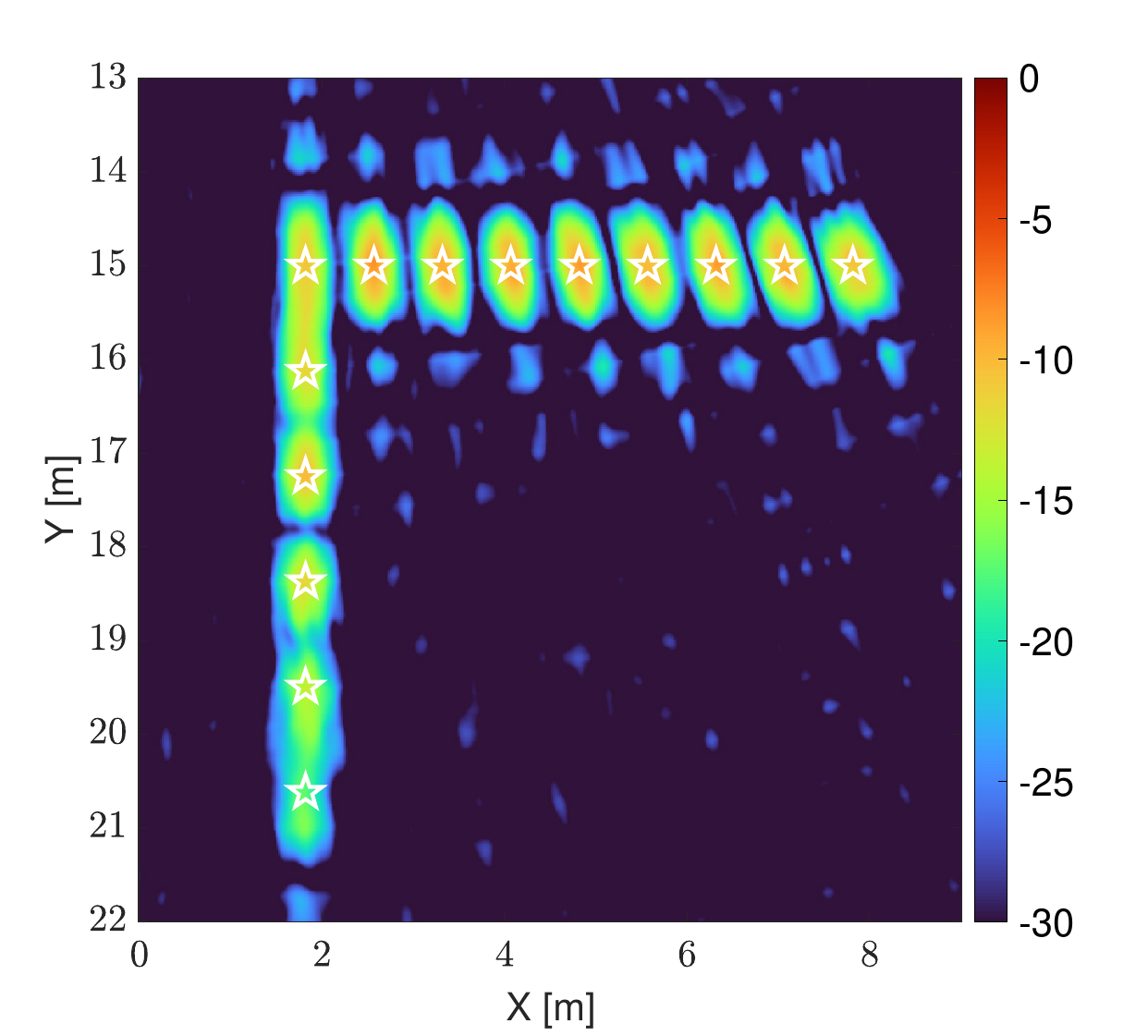}
  }
  \caption{Comparison of images for (a) anomalous mirror with 3GPP standard compliant codebook $\Theta^{\rm 3GPP}_i$; (b) anomalous mirror with proposed codebook $\Theta^{\rm IA}_i$; (c) reflector configured as a lens toward the ROI center, with proposed codebook $\Theta^{\rm IA}_i$; (d) proposed reflector configuration \eqref{eq:angle_periodic_design_quant}, with proposed codebook $\Theta^{\rm IA}_i$. White stars represent the position of the targets. }
  \label{fig:complexcorner}
\end{figure*}

\subsection{Multi-Target Image}

Concerning imaging performance, Fig. \ref{fig:complexcorner} shows a multi-target image in different conditions, for fixed reflector size $A=1.2$ m and $f_0=15$ GHz, aimed at showing the benefits of our proposal w.r.t. benchmarks. We consider 17 point targets with an RCS of $0.01$ m$^2$ deployed along two lines, each exhibiting a scattering phase uncorrelated with all the others\footnote{This assumption was adopted to isolate the effects of the reflection plane configuration itself, without the variability introduced by the target-dependent scattering effects, and it is widely adopted by recent literature \cite{zhi2025nearfieldintegratedimagingcommunication}.}. Fig. \ref{fig:complexcorner}a shows the coherent image obtained by an anomalous mirror and the 3GPP codebook $\Theta^{\rm 3GPP}_i$. In this case, the BS illuminates the reflector with $\overline{K}=3$ orthogonal beams; the sampling condition \eqref{eq:deltathetai} is violated and the image shows evident grating lobes along azimuth. This is the main limitation of NLOS imaging with conventional IA codebooks and it represents the benchmark result a BS would obtain by leveraging the reflection off a building wall. Moreover, targets along $y$ are not illuminated by the mirror, thus they do not appear in the image. The grating lobe issue is solved by using our proposed codebook (Fig. \ref{fig:complexcorner}b), but the reflector used as a mirror only allows to detect 6 out of 17 targets, i.e. the coverage is insufficient. A similar result is obtained by our codebook and a lens toward the center of the ROI, shown in Fig. \ref{fig:complexcorner}c. We notice a resolution enhancement compared to Fig. \ref{fig:complexcorner}b (owing to $A_{\rm eff} = A$ for a lens, while $A_{\rm eff} < A$ for a mirror) but again only a subset of targets are illuminated. Our proposed joint design of BS codebook and reflector modules, whose image is shown in Fig. \ref{fig:complexcorner}d, achieves the best result. Here, we have $N=15$ modules, configured according to \eqref{eq:angle_periodic_design_quant}. At the price of an amplitude/SNR reduction compared to the lens case ($-10$ dB), \textit{all} targets are visible in the image and, mostly, they are imaged at the \textit{same} spatial resolution and with the same amplitude (for reference, the maximum amplitude difference among targets amounts to 5 dB, against 30 dB for the lens or mirror). This confirms the validity of our NLOS imaging approach.

\begin{figure}[!b]
    \centering
    \subfloat[][Setup]{\includegraphics[width=0.9\linewidth]{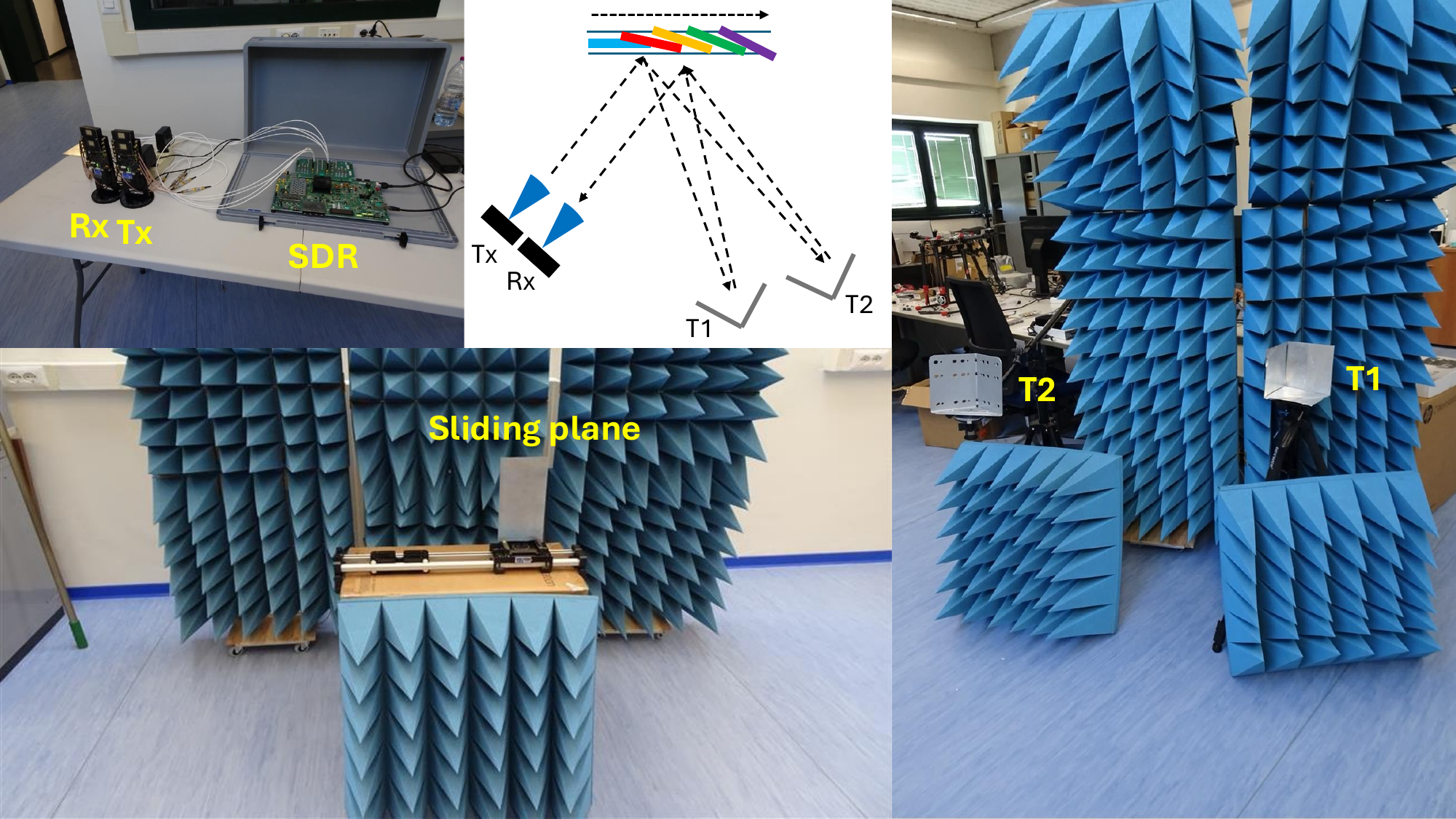}}
        \vspace{-0.3cm} \\
    \subfloat[][Single image]{\includegraphics[width=0.5\linewidth]{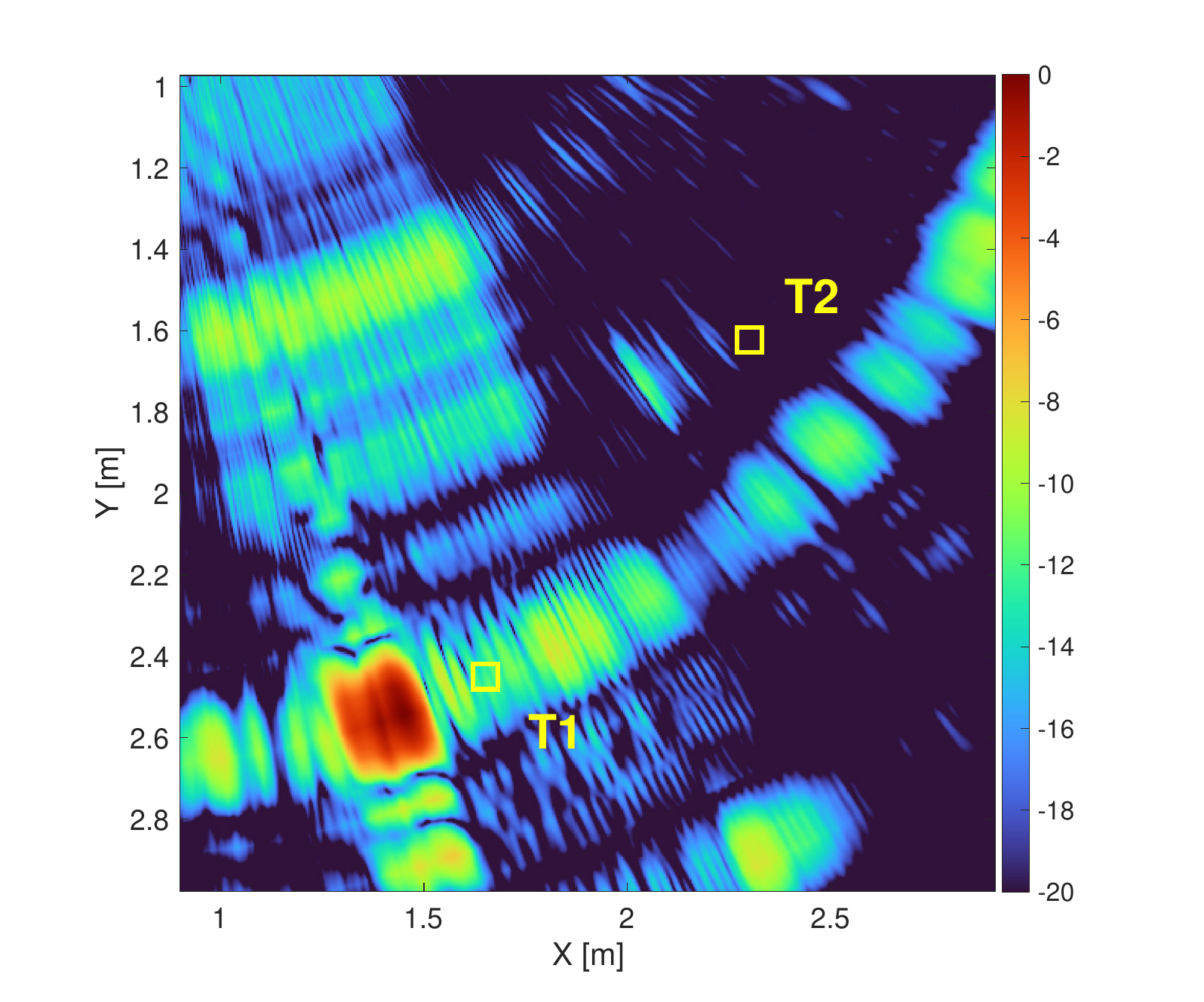}}
            \subfloat[][Coherent image]{\includegraphics[width=0.5\linewidth]{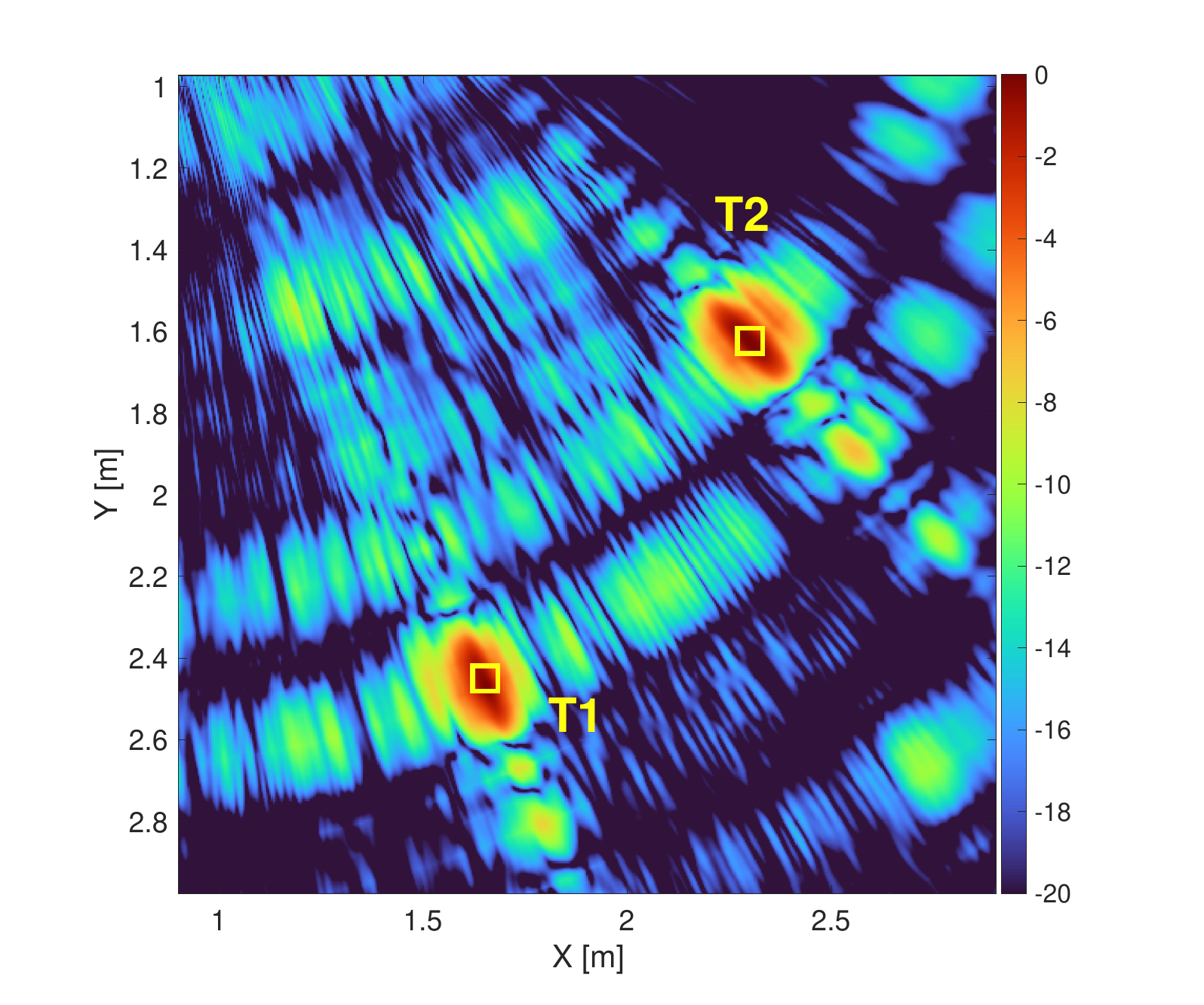}}
    \caption{Experimental setup and imaging results: (a) setup with sliding mirror; (b) image obtained with a single mirror orientation; (c) image obtained by coherently fusing single-orientation images. }
    \label{fig:experiment}
\end{figure}

\subsection{Experimental Evaluation}

The last result showcases a dedicated in-lab experimental test that demonstrates the feasibility and benefits of our imaging method. We implemented a prototype using a Xilinx RFSoC ZCU111~\cite{RFSOC} to generate the baseband transmitted signal and collect the received echoes, interfaced with two Sivers EVK06003 devices \cite{Sivers} (one for Tx and one for Rx, mimicking the BS settings), which were used for conversion of the baseband signal at $f_0 = 60.48$ GHz. Each module includes an RF front end with 16 transmitting and receiving antennas, operating in half-duplex due to thermal constraints. Analog beamforming is performed by the $2\times8$ antenna configuration, over an azimuth beamwidth of $10$ deg. The baseband signal is here a chirp of $B=1024$ MHz bandwidth. The beam is kept fixed and continuously illuminates a planar metallic mirror of $5 \times 20$ cm$^2$, mounted on a motorized rail and moving at constant speed, thereby emulating an aperture of size $A=0.5$ m. Two corner reflectors are placed as in Fig. \ref{fig:experiment}a as targets. To implement the discrete reflection angle pattern \eqref{eq:angle_periodic_design_quant}, we gathered the Rx signal over multiple orientation angles of the metallic mirror. Then, we generate the final image by combining the images at each mirror angle, as shown in Fig. \ref{fig:experiment}a. This approach was adopted to mimic the intended imaging process: since a dense angular beam-sweeping capability is not yet available, the center of the moving mirror was treated as the reference first-order reflection point $\mathbf{p}_\ell$. The objective was to detect their respective contributions by tilting the mirror ---i.e., altering the reflection angle as would be possible with a metasurface--- and coherently combining multiple acquired apertures, in order to enhance both localization accuracy \textit{and} image resolution. The positions of the mirror along the rail w.r.t. the Tx and Rx sources, as well as corner reflector, were calibrated before acquisition. Fig. \ref{fig:experiment}b shows the image obtained by one orientation or the mirror. Out of the two targets (yellow squares), only one is detectable, as predicted by simulations (Fig. \ref{fig:complexcorner}b). Moreover, the detected peak does not correspond to the corner position; this is likely due to a residual sub-centimeter calibration error in the orientation of the rail, that is corrected in post-processing by knowing the ground truth position of corner reflectors. Finally, after multiple acquisitions, the coherent image is formed as in Fig. \ref{fig:experiment}c. Here, calibration residuals were corrected and both targets appear at higher resolution and less angular grating lobes compared to Fig. \ref{fig:complexcorner}b, showing the benefits of our proposal.

\section{Conclusion}
\label{sec:conclusions}
This work proposes and discusses the integration of NLOS imaging into 6G IA procedure. By combining a non-reconfigurable modular reflector with a purposely extended BS beam-sweep codebook, the proposed framework enables high-resolution imaging functionality with limited added overhead in standard IA. Analytical derivations for NF resolution and effective aperture were provided, together with an initial treatment of target motion effects. Simulation results quantified the main trade-offs between imaging performance and IA duration, while experimental tests confirmed the feasibility of the concept. Overall, the results confirmed that embedding ISAC capabilities into the IA process is both achievable and beneficial for future perception-aware 6G networks, whose full hardware integration with communication functions will be explored in future works.

\appendices 

\section{Theoretical vs. Effective Image Resolution}\label{app:resolution}

The resolution of the proposed NLOS imaging system is evaluated invoking diffraction tomography theory (DTT)~\cite{manzoni2023wavefield}. Recall the Rx signal model in Section III, where a source $\mathbf{x}_{\rm bs}$ illuminates a target in $\mathbf{r}$ via double reflection off an anomalous reflector, using a single tone of frequency $f = f_0 + f'$. Neglecting energy losses and noise (irrelevant herein), one can write the Rx signal approximating the scattering from the discrete set of meta-atoms with the scattering from a continuous plane with arbitrary phase configuration.
\begin{figure*}[!t]
\begin{equation}\label{eq:Rxsignal_exttarget_RIS_doublebounce}
\begin{split}
     y(\mathbf{s}, \mathbf{r},f)  & \approx \iint_{S}  \alpha(\mathbf{r}') \int_{x=-\frac{A}{2}}^{\frac{A}{2}} \int_{x'=-\frac{A}{2}}^{\frac{A}{2}} e^{j\phi(x)}e^{j\phi(x')} e^{-jk D_{\mathbf{s}}(x)} e^{-jk D_{\mathbf{r}'}(x)} e^{-jk D_{\mathbf{r}'}(x')} e^{-jk D_{\mathbf{s}}(x')} \mathrm{d}{x}\mathrm{d}x' \mathrm{d}\mathbf{r'}\\
     & \overset{(a)}{\approx}\underbrace{ \int_{x=-\frac{A}{2}}^{\frac{A}{2}} \int_{x'=-\frac{A}{2}}^{\frac{A}{2}}
     e^{j\phi(x)}e^{j\phi(x')} e^{-jk D_{\mathbf{s}}(x)} e^{-jk D_{\mathbf{r}}(x)} e^{-jk D_{\mathbf{r}}(x')} e^{-jk D_{\mathbf{s}}(x')}}_{\text{Reflection gain $G(x)$}}  \iint_{S}  \alpha(\mathbf{r}' ) e^{-j k \left(\nabla D_{ \mathbf{r}}(x) + \nabla D_{ \mathbf{r}}(x')\right)^T \mathbf{r'}}\mathrm{d}\mathbf{r'} \mathrm{d}{x}\mathrm{d}x'
\end{split}
\end{equation}\hrulefill
\end{figure*}
In \eqref{eq:Rxsignal_exttarget_RIS_doublebounce}, \textit{(i)} $S$ is the surface of the target, that is illuminated by the Tx signal, \textit{(ii)} $k = k_0 + k' = 2\pi f/c$ is the spatial angular frequency, \textit{(iii)} $\alpha(\mathbf{r}) = \sqrt{\sigma_\mathbf{r}}e^{j\psi}$ is the target's complex reflectivity, \textit{(iv)} $D_{\mathbf{s}}(x)=\|\mathbf{x}-\mathbf{s}\|$, $D_{\mathbf{r}'}(x)=\|\mathbf{r}'-\mathbf{x}\|$, $D_{\mathbf{r}'}(x')=\|\mathbf{x}'-\mathbf{r}'\|$, $D_{\mathbf{s}}(x')=\|\mathbf{s}-\mathbf{x}'\|$ are the involved distances to/from source and target and points $\mathbf{x}=[x,0]^T$, $\mathbf{x}'=[x',0]^T$. The Rx signal is the sum of all the contributions from each infinitesimal element of the plane and from each infinitesimal point on the target's surface. Approximation $(a)$ is for a Taylor expansion of distances around $\mathbf{r}$, where $\nabla D_{ \mathbf{r}}(x) = (\mathbf{r}-\mathbf{x})/\| \mathbf{r}-\mathbf{x}\|$ and $\nabla D_{\mathbf{r}'}(x') = \mathbf{x}'-\mathbf{r}/ \|\mathbf{x}'-\mathbf{r} \|$ denote unit vectors pointing from position $\mathbf{x}$ on the reflector to the target and from the target to $\mathbf{x}'$, respectively. 
The last integral in \eqref{eq:Rxsignal_exttarget_RIS_doublebounce}$(a)$ is the Fourier transform of the target's reflectivity evaluated in $\mathbf{k}(x,x') = \mathbf{k}_{\rm inc}(x) - \mathbf{k}_{\rm refl}(x')$, where
\begin{align}
   & \mathbf{k}_{\rm inc}(x) = \frac{2 \pi (f_0\hspace{-0.05cm}+\hspace{-0.05cm}f')}{c} \frac{\mathbf{r}-\mathbf{x}}{\| \mathbf{r}-\mathbf{x}\|}, \\
   & \mathbf{k}_{\rm refl}(x') = -\frac{2 \pi (f_0\hspace{-0.05cm}+\hspace{-0.05cm} f')}{c} \frac{\mathbf{x}'-\mathbf{r}}{\| \mathbf{x}'-\mathbf{r}\|}
\end{align}
denote plane wave vectors from the metasurface to the target and vice-versa, respectively. Neglecting multiple bounces as we do in the paper implies that $x=x'$ and $\mathbf{k}(x) = 2 \mathbf{k}_{\rm inc}(x)$. The integration over the reflector area $A$ gives the entire set of illuminated wavevectors, named \textit{spectral coverage}:
\begin{equation}\label{eq:wavenumber_cov_single}
    \mathcal{K}(\mathbf{s},\mathbf{r},A,B) =  \left\{ \mathbf{k}(x) \bigg \lvert x \in \left[-\frac{A}{2},\frac{A}{2}\right], f' \in \left[-\frac{B}{2},\frac{B}{2}\right]\right\}.
\end{equation}
In practice, due to the reflection gain $G(x)$ that is non-constant along the reflector (as discussed in the paper), the spectral coverage is determined by the effective aperture, $\mathcal{K}_{\rm eff}(\mathbf{s},\mathbf{s},A_{\rm eff},B)$. Then, the spatial ambiguity function (SAF) is the image of a point target located in $\mathbf{r}$, and it is defined as
\begin{equation}
    \chi[\mathbf{x}] = \iint_{\mathbf{k} \in (\mathbf{s},\mathbf{s},A_{\rm eff},B)} e^{j \mathbf{k}^T \mathbf{x}} \, \mathrm{d}\mathbf{k}.
\end{equation}
The spatial resolution is defined as the width of the main lobe of the image of a point target located in $\mathbf{r}$. The latter can be derived from the spectral coverage as follows:
\begin{equation}\label{eq:2Dresolution}
     \rho_x = \frac{2\pi}{\Delta k_x},\,\,\rho_y = \frac{2\pi}{\Delta k_y}, \,\,\rho_R= \frac{2\pi}{\Delta k_R},\,\,\rho_\psi= \frac{2\pi}{\Delta k_\psi},
\end{equation}
where $\Delta k_x$, $\Delta k_y$, $\Delta k_R$, $\Delta k_\psi$ denote the width of the spectral coverage along $x$, $y$, range $R$ and azimuth $\psi$. Resolution limit \eqref{eq:2Dresolution} is the lower bound on the effective (achievable) resolution in the proposed system, that is only attained when the elements of the plane are configured to \textit{focus} the impinging signal from the source onto the target, i.e., as a lens, $\phi(x) = 2 k_0 (D_{\mathbf{s}}(x) + D_{\mathbf{r}}(x))$. This limit is attained only as an approximation, and over the effective portion of the plane $A_{\rm eff}$, and the final resolution is ruled by the reflector and not the BS.In general, any configuration different from the lens gives a resolution loss. In case the reflector configured to behave as a mirror (i.e., according to a single angle), such that $\phi(x) = \alpha x$, the resolution of the image is dictated by the BS and \textit{not} by the plane. 

\section{Proof of Proposition on NF Resolution} \label{app:proof}

The computation of the NF resolution along range and azimuth for a target in $\mathbf{r}=(R_0,\psi_0)$ is derived from the wavenumber coverage \eqref{eq:wavenumber_cov_single}. Now, we want to compute the resolution in polar coordinates $\rho_R$ and $\rho_\psi$. Let us consider a single frequency $f_0$. The spectral coverage $\mathcal{K}$ is the locus of points defined by the resulting wavevector:
\begin{align}
    \mathbf{k}(x) = \frac{4 \pi f_0}{c} & [\sin \psi(x), \cos \psi(x)]^T \\
    & \text{with} \,\,\psi(x) = \arctan\left(\frac{R_0 \sin \psi_0 - x}{R_0 \cos \psi_0}\right) \nonumber
\end{align}
From $x=-A_{\rm eff}/2$ to $x=A_{\rm eff}/2$, the spectral coverage in NF forms an annular region in the 2D spatial frequency domain, while it degenerates to a point in FF. 

The width of the spectral coverage $\Delta k_R$, dictating the range resolution, is the distance from the arc to the chord formed by the line connecting the ending points $x=-A_{\rm eff}/2$ and $x=A_{\rm eff}/2$. By basic geometric derivations, we have:
\begin{equation}
\begin{split}
    \Delta k_R & = \frac{4 \pi f_0}{c} \left[ 1 - \cos\left(\mathsf{F}_+\right)\right]
\end{split}
\end{equation}
and the term inside the cosine is the $\mathsf{F}_+= \psi(x=A_{\rm eff}/2)-\psi_0$ factor. In NF, due to the curvature of the wavefront, we have $\Delta k_R > 0$ and thus a finite range resolution $\rho_R$ even with a single frequency. In FF, $A_{\rm eff}/R_0 \rightarrow 0$, thus $\psi(x=A_{\rm eff}/2) \rightarrow \psi_0$ and $\Delta k_R \rightarrow 0$. In this latter case, we do not have range resolution. For a finite bandwidth $B$, factor $\Delta k_R$ has an additional term equal to $4 \pi B/c$, and we can match the expression in proposition by introducing the bandwidth augmentation factor $\kappa_R = \Delta k_R/(4 \pi B/c)$. 

The width of the spectral coverage along azimuth, $\Delta k_\psi$, is related to the spectral coverage in cross-range $\Delta k_{XR}$, that is in turn determined by the length of the aforementioned chord:
\begin{equation}
    \Delta k_{XR} = \frac{4 \pi f_0}{c} \left[\sin\left(\mathsf{F}_+\right) - \sin\left(\mathsf{F}_-\right)\right]
\end{equation}
where $\mathsf{F}_- =  \psi(x=-A_{\rm eff}/2)-\psi_0$. The cross-range resolution is $\rho_{XR} = 2\pi/\Delta k_{XR}$ and the azimuth resolution follows after $\rho_{\psi} = \rho_{XR}/R_0$. By equating $\rho_{\psi}$ with the NF expression in proposition, we derive the effective area factor $\kappa_\psi$.

\bibliographystyle{IEEEtran}
\bibliography{Bibliography,Bibliography_TWC,bibliography_2}

\begin{thebibliography}{10}
\providecommand{\url}[1]{#1}
\csname url@samestyle\endcsname
\providecommand{\newblock}{\relax}
\providecommand{\bibinfo}[2]{#2}
\providecommand{\BIBentrySTDinterwordspacing}{\spaceskip=0pt\relax}
\providecommand{\BIBentryALTinterwordstretchfactor}{4}
\providecommand{\BIBentryALTinterwordspacing}{\spaceskip=\fontdimen2\font plus
\BIBentryALTinterwordstretchfactor\fontdimen3\font minus
  \fontdimen4\font\relax}
\providecommand{\BIBforeignlanguage}[2]{{%
\expandafter\ifx\csname l@#1\endcsname\relax
\typeout{** WARNING: IEEEtran.bst: No hyphenation pattern has been}%
\typeout{** loaded for the language `#1'. Using the pattern for}%
\typeout{** the default language instead.}%
\else
\language=\csname l@#1\endcsname
\fi
#2}}
\providecommand{\BIBdecl}{\relax}
\BIBdecl

\bibitem{Gonzalez-Prelcic2025}
N.~Gonzalez-Prelcic, D.~Tagliaferri, M.~F. Keskin, H.~Wymeersch, and L.~Song,
  ``Six integration avenues for isac in 6g and beyond,'' \emph{IEEE Vehicular
  Technology Magazine}, vol.~20, no.~1, pp. 18--39, 2025.

\bibitem{10287134}
S.~E. Trevlakis, A.-A.~A. Boulogeorgos, D.~Pliatsios, J.~Querol, K.~Ntontin,
  P.~Sarigiannidis, S.~Chatzinotas, and M.~D. Renzo, ``Localization as a key
  enabler of 6g wireless systems: A comprehensive survey and an outlook,''
  \emph{IEEE Open Journal of the Communications Society}, pp. 1--1, 2023.

\bibitem{manzoni2023wavefield}
M.~Manzoni, D.~Tagliaferri, S.~Tebaldini, M.~Mizmizi, A.~V. Monti-Guarnieri,
  C.~M. Prati, and U.~Spagnolini, ``Wavefield networked sensing: Principles,
  algorithms and applications,'' pp. 181--197, 2025.

\bibitem{Masouros2025_distributedISAC}
K.~Meng, C.~Masouros, A.~P. Petropulu, and L.~Hanzo, ``Cooperative isac
  networks: Performance analysis, scaling laws, and optimization,'' \emph{IEEE
  Transactions on Wireless Communications}, vol.~24, no.~2, pp. 877--892, 2025.

\bibitem{ManzoniCOSMIC}
M.~Manzoni, F.~Linsalata, M.~Magarini, and S.~Tebaldini, ``Cosmic waveforms for
  integrated communication and imaging,'' in \emph{ICASSP 2025 - 2025 IEEE
  International Conference on Acoustics, Speech and Signal Processing
  (ICASSP)}, 2025, pp. 1--5.

\bibitem{Li2024}
K.~Li, D.~Ramirez, K.~V. Mishra, and A.~Sabharwal, ``Repurposing mu-mimo
  downlink for joint wireless communications and imaging via virtual users,''
  in \emph{ICASSP 2024 - 2024 IEEE International Conference on Acoustics,
  Speech and Signal Processing (ICASSP)}, 2024, pp. 13\,061--13\,065.

\bibitem{11087660}
N.~Abusanad, M.~Ayasrah, S.~Aïssa, and H.~Arslan, ``Joint imaging and downlink
  communication with shared resources,'' \emph{IEEE Transactions on Vehicular
  Technology}, pp. 1--16, 2025.

\bibitem{10097213}
X.~Tong, Z.~Zhang, and Z.~Yang, ``Multi-view millimeter-wave imaging over
  wireless cellular network,'' in \emph{ICASSP 2023 - 2023 IEEE International
  Conference on Acoustics, Speech and Signal Processing (ICASSP)}, 2023, pp.
  1--5.

\bibitem{10540249}
J.~Li, X.~Shao, F.~Chen, S.~Wan, C.~Liu, Z.~Wei, and D.~Wing Kwan~Ng,
  ``Networked integrated sensing and communications for 6g wireless systems,''
  \emph{IEEE Internet of Things Journal}, vol.~11, no.~17, pp.
  29\,062--29\,075, 2024.

\bibitem{IIAC_3D_imaging}
H.~S. Rou, G.~T.~F. de~Abreu, D.~Gonzalez~G., and O.~Gonsa, ``Integrated
  sensing and communications for 3d object imaging via bilinear inference,''
  \emph{IEEE Transactions on Wireless Communications}, vol.~23, no.~8, pp.
  8636--8653, 2024.

\bibitem{zhi2025nearfieldintegratedimagingcommunication}
\BIBentryALTinterwordspacing
K.~Zhi, T.~Yang, S.~Li, Y.~Song, A.~Rezaei, and G.~Caire, ``Near-field
  integrated imaging and communication in distributed mimo networks,'' 2025.
  [Online]. Available: \url{https://arxiv.org/abs/2508.17526}
\BIBentrySTDinterwordspacing

\bibitem{6798744}
L.~Lu, G.~Y. Li, A.~L. Swindlehurst, A.~Ashikhmin, and R.~Zhang, ``An overview
  of massive mimo: Benefits and challenges,'' \emph{IEEE Journal of Selected
  Topics in Signal Processing}, vol.~8, no.~5, pp. 742--758, 2014.

\bibitem{9468353}
D.~Solomitckii, M.~Heino, S.~Buddappagari, M.~A. Hein, and M.~Valkama, ``Radar
  scheme with raised reflector for nlos vehicle detection,'' \emph{IEEE
  Transactions on Intelligent Transportation Systems}, vol.~23, no.~7, pp.
  9037--9045, 2022.

\bibitem{9553059}
J.~Wei, S.~Wei, X.~Liu, M.~Wang, J.~Shi, and X.~Zhang, ``Non-line-of-sight
  imaging by millimeter wave radar,'' in \emph{2021 IEEE International
  Geoscience and Remote Sensing Symposium IGARSS}, 2021, pp. 2983--2986.

\bibitem{9547412}
S.~Wei, J.~Wei, X.~Liu, M.~Wang, S.~Liu, F.~Fan, X.~Zhang, J.~Shi, and G.~Cui,
  ``Nonline-of-sight 3-d imaging using millimeter-wave radar,'' \emph{IEEE
  Transactions on Geoscience and Remote Sensing}, vol.~60, pp. 1--18, 2022.

\bibitem{11065141}
Y.~Zhou, T.~Yang, and D.~He, ``Nlos localization method based on kirchhoff
  migration,'' in \emph{2025 6th International Conference on Geology, Mapping
  and Remote Sensing (ICGMRS)}, 2025, pp. 581--585.

\bibitem{7362138}
M.~Gustafsson, A.~Andersson, T.~Johansson, S.~Nilsson, A.~Sume, and A.~Orbom,
  ``Extraction of human micro-doppler signature in an urban environment using a
  “sensing-behind-the-corner” radar,'' \emph{IEEE Geoscience and Remote
  Sensing Letters}, vol.~13, no.~2, pp. 187--191, 2016.

\bibitem{8966246}
S.~Guo, Q.~Zhao, G.~Cui, S.~Li, L.~Kong, and X.~Yang, ``Behind corner targets
  location using small aperture millimeter wave radar in nlos urban
  environment,'' \emph{IEEE Journal of Selected Topics in Applied Earth
  Observations and Remote Sensing}, vol.~13, pp. 460--470, 2020.

\bibitem{10298635}
Y.~Xu, G.~Liu, and T.~Jiang, ``Leveraging rough-relay-surface scattering for
  non-line-of-sight mmwave radar sensing,'' \emph{IEEE Internet of Things
  Journal}, vol.~11, no.~6, pp. 10\,964--10\,978, 2024.

\bibitem{7109827}
G.~Oliveri, D.~H. Werner, and A.~Massa, ``Reconfigurable electromagnetics
  through metamaterials—a review,'' \emph{Proceedings of the IEEE}, vol. 103,
  no.~7, pp. 1034--1056, 2015.

\bibitem{Buzzi_RISforradar_journal}
S.~Buzzi, E.~Grossi, M.~Lops, and L.~Venturino, ``{Foundations of MIMO Radar
  Detection Aided by Reconfigurable Intelligent Surfaces},'' \emph{IEEE
  Transactions on Signal Processing}, vol.~70, pp. 1749--1763, 2022.

\bibitem{10753053}
J.~Wang, J.~Fang, H.~Li, and L.~Huang, ``Intelligent reflecting
  surface-assisted nlos sensing with ofdm signals,'' \emph{IEEE Transactions on
  Signal Processing}, vol.~72, pp. 5322--5337, 2024.

\bibitem{9299878}
Y.~Tao and Z.~Zhang, ``Distributed computational imaging with reconfigurable
  intelligent surface,'' in \emph{2020 International Conference on Wireless
  Communications and Signal Processing (WCSP)}, 2020, pp. 448--454.

\bibitem{jiang2023near}
Y.~Jiang, F.~Gao, Y.~Liu, S.~Jin, and T.~Cui, ``Near field computational
  imaging with ris generated virtual masks,'' 2023.

\bibitem{torcolacci2023holographic}
G.~Torcolacci, A.~Guerra, H.~Zhang, F.~Guidi, Q.~Yang, Y.~C. Eldar, and
  D.~Dardari, ``{Holographic Imaging with XL-MIMO and RIS: Illumination and
  Reflection Design},'' \emph{IEEE Journal of Selected Topics in Signal
  Processing}, pp. 1--16, 2024.

\bibitem{10526279}
H.~Sun, F.~Gao, S.~Zhang, S.~Jin, and T.~Jun~Cui, ``Computational imaging with
  holographic ris: Sensing principle and pathloss analysis,'' \emph{IEEE
  Journal on Selected Areas in Communications}, vol.~42, no.~6, pp. 1703--1716,
  2024.

\bibitem{9650562}
N.~Ghavami, E.~Razzicchia, O.~Karadima, P.~Lu, W.~Guo, I.~Sotiriou, E.~Kallos,
  G.~Palikaras, and P.~Kosmas, ``The use of metasurfaces to enhance microwave
  imaging: Experimental validation for tomographic and radar-based
  algorithms,'' \emph{IEEE Open Journal of Antennas and Propagation}, vol.~3,
  pp. 89--100, 2022.

\bibitem{10618967}
Z.~Li, A.~Dubey, S.~Shen, N.~K. Kundu, J.~Rao, and R.~Murch, ``Radio
  tomographic imaging with reconfigurable intelligent surfaces,'' \emph{IEEE
  Transactions on Wireless Communications}, vol.~23, no.~11, pp.
  15\,784--15\,797, 2024.

\bibitem{10694426}
P.~Tosi, M.~Henninger, L.~G. de~Oliveira, and S.~Mandelli, ``Feasibility of
  non-line-of-sight integrated sensing and communication at mmwave,'' in
  \emph{2024 IEEE 25th International Workshop on Signal Processing Advances in
  Wireless Communications (SPAWC)}, 2024, pp. 331--335.

\bibitem{jsan13010002}
\BIBentryALTinterwordspacing
T.~Althobaiti, R.~A. Khalil, and N.~Saeed, ``Robust isac localization in smart
  cities: A hybrid network approach for nlos challenges with uncertain
  parameters,'' \emph{Journal of Sensor and Actuator Networks}, vol.~13, no.~1,
  2024. [Online]. Available: \url{https://www.mdpi.com/2224-2708/13/1/2}
\BIBentrySTDinterwordspacing

\bibitem{Khosroshahi2024Localization}
\BIBentryALTinterwordspacing
K.~Khosroshahi, P.~Sehier, S.~Mekki, and M.~Suppa, ``Localization accuracy
  improvement in multistatic isac with los/nlos condition using 5g nr
  signals,'' \emph{arXiv preprint}, no. arXiv:2412.17577, 2024. [Online].
  Available: \url{https://arxiv.org/abs/2412.17577}
\BIBentrySTDinterwordspacing

\bibitem{Paglierani2025_DT_localization}
N.~Paglierani, F.~Linsalata, O.~A. Sevimay, L.~Cazzella, D.~Badini,
  M.~Magarini, and U.~Spagnolini, ``Digital network twin-enabled
  synchronization and localization,'' \emph{IEEE Journal on Selected Areas in
  Communications}, pp. 1--1, 2025.

\bibitem{bellini_2025JCNS}
D.~T. Bellini, D.~Tagliaferri, M.~Mizmizi, S.~Tebaldini, and U.~Spagnolini,
  ``Multi-view integrated imaging and communication,'' in \emph{2025 IEEE 5th
  International Symposium on Joint Communications \& Sensing (JC\&S)}, 2025,
  pp. 1--6.

\bibitem{bellini2024multiview}
D.~Tornielli~Bellini, D.~Tagliaferri, M.~Mizmizi, S.~Tebaldini, and
  U.~Spagnolini, ``Multi-view near-field imaging in nlos with
  non-reconfigurable em skins,'' in \emph{2024 IEEE International Conference on
  Communications Workshops (ICC Workshops)}, 2024, pp. 384--389.

\bibitem{giordani2018tutorial}
M.~Giordani, M.~Polese, A.~Roy, D.~Castor, and M.~Zorzi, ``A tutorial on beam
  management for 3gpp nr at mmwave frequencies,'' \emph{IEEE Communications
  Surveys \& Tutorials}, vol.~21, no.~1, pp. 173--196, 2018.

\bibitem{1281676}
\BIBentryALTinterwordspacing
T.~Ropitault, S.~Blandino, D.~Griffith, T.~Nguyen, A.~Sahoo, and N.~Golmie,
  ``\BIBforeignlanguage{en}{Characterization of monostatic base stations
  sensing resolution using 5g reference signals}.''\hskip 1em plus 0.5em minus
  0.4em\relax IEEE International Conference on Communication (ICC) WS17:
  Workshop on Positioning and Sensing Over Wireless Networks , Montreal, CA,
  2025-07-01 04:07:00 2025. [Online]. Available:
  \url{https://tsapps.nist.gov/publication/get_pdf.cfm?pub_id=958896}
\BIBentrySTDinterwordspacing

\bibitem{10083170}
K.~Abratkiewicz, A.~Ksiezyk, M.~Płotka, P.~Samczynski, J.~Wszołek, and T.~P.
  Zielinski, ``Ssb-based signal processing for passive radar using a 5g
  network,'' \emph{IEEE Journal of Selected Topics in Applied Earth
  Observations and Remote Sensing}, vol.~16, pp. 3469--3484, 2023.

\bibitem{10200933}
M.~Golzadeh, E.~Tiirola, L.~Anttila, J.~Talvitie, K.~Hooli, O.~Tervo,
  I.~Peruga, S.~Hakola, and M.~Valkama, ``Downlink sensing in 5g-advanced and
  6g:sib1-assisted ssb approach,'' in \emph{2023 IEEE 97th Vehicular Technology
  Conference (VTC2023-Spring)}, 2023, pp. 1--7.

\bibitem{10502156}
Y.~Li, F.~Liu, Z.~Du, W.~Yuan, Q.~Shi, and C.~Masouros, ``Frame structure and
  protocol design for sensing-assisted nr-v2x communications,'' \emph{IEEE
  Transactions on Mobile Computing}, vol.~23, no.~12, pp. 11\,045--11\,060,
  2024.

\bibitem{golzadeh2024prs_ambiguity}
M.~Golzadeh, E.~Tiirola, J.~Talvitie, L.~Anttila, K.~Hooli, O.~Tervo, and
  M.~Valkama, ``{Joint Sensing and UE Positioning in 5G-6G: PRS Range
  Estimation with Suppressed Ambiguity},'' in \emph{Proceedings of the IEEE
  Radar Conference (RadarConf) 2024}.\hskip 1em plus 0.5em minus 0.4em\relax
  IEEE, May 2024.

\bibitem{jopanya2025ssb_uav_detection}
P.~Jopanya and D.~P.~M. Osorio, ``{Utilizing 5G NR SSB Blocks for Passive
  Detection and Localization of Low-Altitude Drones},'' \emph{arXiv preprint
  arXiv:2504.02641v2}, 2025, updated version (v2) on 12 Apr 2025.

\bibitem{10382696}
Z.~Xiao, S.~Chen, and Y.~Zeng, ``Simultaneous multi-beam sweeping for mmwave
  massive mimo integrated sensing and communication,'' \emph{IEEE Transactions
  on Vehicular Technology}, vol.~73, no.~6, pp. 8141--8152, 2024.

\bibitem{wachowiak2025approximationrangeambiguityfunction}
\BIBentryALTinterwordspacing
M.~Wachowiak, A.~Bourdoux, and S.~Pollin, ``Approximation of the range
  ambiguity function in near-field sensing systems,'' 2025. [Online].
  Available: \url{https://arxiv.org/abs/2509.22423}
\BIBentrySTDinterwordspacing

\bibitem{10621891}
S.~Häger, M.~Danger, K.~Heimann, Y.~Gümüs, S.~Böeker, and C.~Wietfeld,
  ``Custom design and experimental evaluation of passive reflectors for mmwave
  private networks,'' in \emph{2024 IEEE 30th International Symposium on Local
  and Metropolitan Area Networks (LANMAN)}, 2024, pp. 52--57.

\bibitem{10926852}
R.~Aghazadeh~Ayoubi, S.~Mura, D.~Tagliaferri, M.~Mizmizi, and U.~Spagnolini,
  ``Optimizing curved em skins for opportunistic passive reflection in
  vehicular networks,'' \emph{IEEE Transactions on Communications}, vol.~73,
  no.~9, pp. 8001--8015, 2025.

\bibitem{manzoni2022motion}
M.~Manzoni, D.~Tagliaferri, M.~Rizzi, S.~Tebaldini, A.~V. Monti-Guarnieri,
  C.~M. Prati, M.~Nicoli, I.~Russo, S.~Duque, C.~Mazzucco, and U.~Spagnolini,
  ``Motion estimation and compensation in automotive mimo sar,'' pp.
  1756--1772, 2023.

\bibitem{Giovannetti_NF_velocity}
C.~Giovannetti, N.~Decarli, and D.~Dardari, ``Performance bounds for velocity
  estimation with extremely large aperture arrays,'' \emph{IEEE Wireless
  Communications Letters}, vol.~13, no.~12, pp. 3513--3517, 2024.

\bibitem{FrFT}
A.~Serbes and O.~Aldimashki, ``A fast and accurate chirp rate estimation
  algorithm based on the fractional fourier transform,'' in \emph{2017 25th
  European Signal Processing Conference (EUSIPCO)}, 2017, pp. 1105--1109.

\bibitem{Giordani_beammanagement}
M.~Giordani, M.~Polese, A.~Roy, D.~Castor, and M.~Zorzi, ``A tutorial on beam
  management for 3gpp nr at mmwave frequencies,'' \emph{IEEE Communications
  Surveys \& Tutorials}, vol.~21, no.~1, pp. 173--196, 2019.

\bibitem{RFSOC}
AMD, ``Zynq ultrascale+ rfsoc data sheet: Overview,'' 2023, version 1.14.

\bibitem{Sivers}
\BIBentryALTinterwordspacing
S.~Semiconductors. Evaluation kits and evaluation boards. [Online]. Available:
  \url{https://www.sivers-semiconductors.com/sivers-wireless/wireless-products/evaluation-kits/}
\BIBentrySTDinterwordspacing

\end{thebibliography}

\end{document}